\documentclass[aps,prd,reprint,onecolumn,superscriptaddress,nofootinbib]{revtex4}
\pdfoutput=1
\usepackage{epsfig, subfigure}
\usepackage{setspace}
\usepackage{booktabs, tabularx} 
 
\usepackage{amsmath,bm,bbm}
\usepackage{hyperref}
\usepackage{graphicx}
\usepackage{enumerate}
\hypersetup{
    pdfnewwindow=true,      
    colorlinks=true,       
    linkcolor=blue,          
    citecolor=blue,        
    filecolor=blue,      
    urlcolor=blue           
}

\def\lsim{\lesssim}  
\def\gsim{\gtrsim}

\newcommand{\beq}{\begin{equation}}  
\newcommand{\eeq}{\end{equation}}
\newcommand{\bea}{\begin{eqnarray}}  
\newcommand{\eea}{\end{eqnarray}}
\newcommand{\epm}{\ensuremath{e^{\pm}\;}}

\newcommand{\neff}{{\rm N}_{\rm eff}}
\newcommand{\Deln}{\ensuremath{\Delta{\rm N}_\nu}}

\newcommand{\ie}{{\it i.e.}}
\newcommand{\eg}{{\it e.g.}}
\newcommand{\etal}{{\it et al.}}

\begin{document}

\title{\mbox{\hspace{-0.5cm}Equivalent Neutrinos, Light WIMPs, and the Chimera of Dark Radiation}}

\author{Gary Steigman}
\email{steigman.1@asc.ohio-state.edu}
\affiliation{Center for Cosmology and Astro-Particle Physics,\\
Department of Physics, Department of Astronomy,\\
The Ohio State University,  Columbus, OH 43210, USA
}


\date{\today}

\begin{abstract}
According to conventional wisdom, in the standard model (SM) of particle physics and cosmology the ``effective number of neutrinos" measured in the late Universe is $\neff = 3$ (more precisely, 3.046).  For extensions of the standard model allowing for the presence of  \Deln~``equivalent neutrinos" (or ``dark radiation"), it is generally the case that $\neff > 3$.  These canonical results are reconsidered, demonstrating that a measurement of $\neff > 3$ can be consistent with \Deln~= 0 (``dark radiation without dark radiation").  Conversely, a measurement consistent with $\neff = 3$ is not inconsistent with the presence of dark radiation ($\Delta{\rm N}_{\nu} > 0$).  In particular, if there is a light WIMP that annihilates to photons after the SM neutrinos have decoupled, the photons are heated beyond their usual heating from \epm annihilation, reducing the late time ratio of neutrino and photon temperatures (and number densities), leading to $\neff < 3$.  This opens the window for one or more equivalent neutrinos, including ``sterile neutrinos", to be consistent with $\neff = 3$.  The absence of evidence for equivalent neutrinos is not evidence for the absence of equivalent neutrinos.  By reducing the neutrino number density in the present Universe, this also allows for more massive neutrinos, relaxing the current constraints on the sum of the neutrino masses.  In contrast, if the light WIMP couples only to the SM neutrinos and not to the photons and \epm pairs, its late time annihilation heats the neutrinos but not the photons, resulting in $\neff > 3$ even in the absence of equivalent neutrinos or dark radiation.  A measurement of $\neff > 3$ is no guarantee of the presence of equivalent neutrinos or dark radiation.  In the presence of a light WIMP and/or equivalent neutrinos there are degeneracies among the light WIMP mass and its nature (fermion or boson, as well as its couplings to neutrinos and photons), the number and nature (fermion or boson) of the equivalent neutrinos, and their decoupling temperature (the strength of their interactions with the SM particles).  As the analysis here reveals, there's more to a measurement of $\neff$ than meets the eye.
\end{abstract}

\pacs{95.35.+d}
\keywords{Dark Matter}
\maketitle


\section{Introduction}
\label{intro}

For the standard model (SM) of particle physics and cosmology at late times in the early Universe, after the \epm pairs have annihilated, the only massless or extremely relativistic particles remaining are the photons and the three SM neutrinos.  In the SM the neutrinos decouple prior to \epm annihilation so that only the photons are heated when the pairs annihilate.  The strength of the SM weak interactions determines the neutrino decoupling temperature which, in turn, fixes the relative contributions of the photons and neutrinos to the late time, early Universe (radiation dominated) energy density.  This relative contribution of neutrinos, measured by the ``effective number of neutrinos" is $\neff = 3$ under the assumptions of the standard models of particle physics and cosmology.  Many years ago, stimulated by the desire to test the prediction of asymptotic freedom limiting the number of particle physics families \cite{gross,politzer}, and by the discovery of the third family of leptons, along with its neutrino \cite{perl}, which led to $\neff$ increasing from 2 to 3, Steigman, Schramm, and Gunn \cite{ssg} explored the consequences for big bang nucleosynthesis (BBN) of additional, ``equivalent neutrinos" (see, also, the earlier related work of Hoyle and Tayler \cite{hoyle}, of Peebles \cite{pje} and of Shvartsman \cite{shvartsman}).  Ever since it has been a goal of a broad array of cosmological observations, from those of the abundances of the light elements produced during BBN to studies of the cosmic microwave background (CMB) radiation and of large scale structure (LSS), to measure $\neff$.  In recent years both BBN and the CMB/LSS have favored values of $\neff > 3$ \cite{steigman,komatsu,dunkley,keisler,archidiacono}, hinting at the presence of equivalent neutrinos or ``dark radiation".  In anticipation that the results from the Planck experiment \cite{galli} will provide the most accurate determination of $\neff$ to date, it is timely to revisit the theoretical predictions for models beyond the SM containing equivalent neutrinos and WIMPs, weakly interacting massive particles that are candidates for the dark matter in the Universe.  In the course of the analysis presented here, several degeneracies are noted in the determination of $\neff$ that will render the interpretation of any precision measurement of $\neff$ more problematic, and more interesting.

In \S\,\ref{sm} the standard model analysis is reviewed, allowing the neutrino decoupling temperature ($T_{\nu d}$) to be a free parameter, revealing how $\neff$ depends on its value.  In the process, very small differences with the canonical, textbook results are revealed.  With this as background, in \S\,\ref{equivnus} the analysis is extended to allow for equivalent neutrinos ($\xi$).  It is noted here that $\neff$ now depends on the number (\Deln) and nature (fermion or boson) of the equivalent neutrinos as well as on their decoupling temperature (how strongly they couple to the SM particles).  Sterile neutrinos  are equivalent neutrinos (Majorana fermions) that decouple along with the SM neutrinos ($T_{\xi d} = T_{\nu d}$), but more general equivalent neutrinos may decouple before or after the SM neutrinos ($T_{\xi d} \neq T_{\nu d}$), affecting both $\neff$ and the connection between the sum of the neutrino masses and their contribution to the present Universe mass density (for neutrinos with non-zero mass).  This analysis is further extended in \S\,\ref{light wimp} to allow for the presence of light WIMPs ($\chi$) whose annihilation occurs around or after the time when the SM neutrinos decouple.  The light WIMP annihilation can heat the photons beyond the usual heating from \epm annihilation, reducing the relative contribution of the neutrinos to the energy density, leading to $\neff < 3$.  In this case the degeneracies discussed above are expanded to include the nature (fermion or boson) of the WIMP and its mass ($m_{\chi}$).  Many more possibilities now emerge allowing, for example, for one or even two sterile neutrinos (\Deln~= 1, 2) even if observations should find $\neff \lsim 3.5$.  It is shown that an observation of $\neff = 3$ would not exclude the presence of equivalent neutrinos or dark radiation.  The tables are turned in \S\,\ref{weak wimp} where it is assumed that the light WIMP couples only to the SM neutrinos and not to the photons or \epm pairs.  In this case the SM neutrinos are heated by WIMP annihilation, increasing their relative contribution to the early Universe energy density, resulting in $\neff > 3$ even in the absence of equivalent neutrinos or dark radiation.  The results are reviewed and summarized in \S\,\ref{summary}.

\section {Standard Model Neutrinos}
\label{sm}

To set the stage for the subsequent discussion, in this section it is assumed that there are no light WIMPs (\eg, with $m_{\chi}\lsim 20\,{\rm MeV}$) or ``extra" neutrinos (\eg, sterile neutrinos) or other relativistic particles (equivalent neutrinos), only the standard model particles including the three SM neutrinos.  However, the neutrino decoupling temperature, $T_{\nu d}$, assumed to be the same for all three flavors, is allowed to be a free parameter.  Allowing $T_{\nu d}$ to vary is equivalent to imagining that the weak interactions are weaker, or stronger, than the SM weak interactions.  Of course, the strength of the weak interactions and $T_{\nu d}$ are determined by laboratory and accelerator experiments ($T_{\nu d} \approx 2 - 3\,{\rm MeV}$ \cite{enqvist,dolgov,hannestad}) and $T_{\nu d}$ is not really a free parameter.  However, it is interesting and informative to ask, ``How does allowing the neutrino decoupling temperature to vary change the well known, canonical SM neutrino results?".  To facilitate comparison with the usual SM results, the neutrinos are assumed to decouple instantaneously, when $T_{\gamma} = T_{\nu d}$.  For the analysis here, the instantaneous decoupling approximation, typically accurate to $\sim 2\,\%$ or better, replaces a coupled set of integro-differential equations which need to be solved numerically (see, \eg,\,\cite{dolgov,mangano}), with algebraic equations that follow from entropy conservation.  

Prior to neutrino decoupling, for $T_{\gamma} \geq T_{\nu d}$\,, $T_{\nu} = T_{\gamma}$.  After neutrino decoupling, for $T_{\gamma} < T_{\nu d}$\,, $T_{\nu} \leq T_{\gamma}$ as a consequence of the heating of the photons relative to the decoupled neutrinos.  After \epm annihilation, when $T_{\gamma} = T_{\gamma 0}$, where $T_{\gamma 0} \ll {\rm min}\{m_{e},T_{\nu d}\}$, entropy conservation permits the calculation of the ``frozen out" ratio of neutrino and photon temperatures (and number densities).  The result is,
\beq
\bigg({T_{\nu} \over T_{\gamma}}\bigg)^{3}_{0} = {g_{\gamma} \over g_{s}(T_{\nu d}) - 3g_{\nu}} = {2 \over g_{s}(T_{\nu d}) - 21/4}\,,
\eeq
where $g_{s} = g_{s}(T)$ is defined by the ratio of the total entropy density to the entropy density contributed by photons alone,
\beq
s_{tot}/s_{\gamma} \equiv g_{s}/g_{\gamma} = g_{s}/2\,,
\eeq
and the entropy density at temperature $T$ is defined by,
\beq
s \equiv {\rho + \it{p} \over T}\,,
\eeq
where $\rho$ is the energy density and $\it{p}$ is the pressure.  As a result,
\beq
{11 \over 4}\bigg({T_{\nu} \over T_{\gamma}}\bigg)^{3}_{0} = {11 \over 2\,g_{s}(T_{\nu d}) - 10.5}\,.
\eeq

In the canonical, textbook analysis it is {\it assumed} that the neutrinos decouple instantaneously, and that at neutrino decoupling only the photons, the \epm pairs, and the three SM neutrinos contribute to $g_{s}$.  It is a further, unstated assumption that at neutrino decoupling the \epm pairs are essentially massless so that $(s_{e}/s_{\gamma})_{T_{\nu d}} = 7/4$.  That is, it is {\it assumed} that $g_{s}(T_{\nu d}) = 43/4 = 10.75$, resulting in $(T_{\nu}/T_{\gamma})^{3}_{0} = 4/11$.  

\begin{figure}[!t]
\includegraphics[width=0.60\columnwidth]{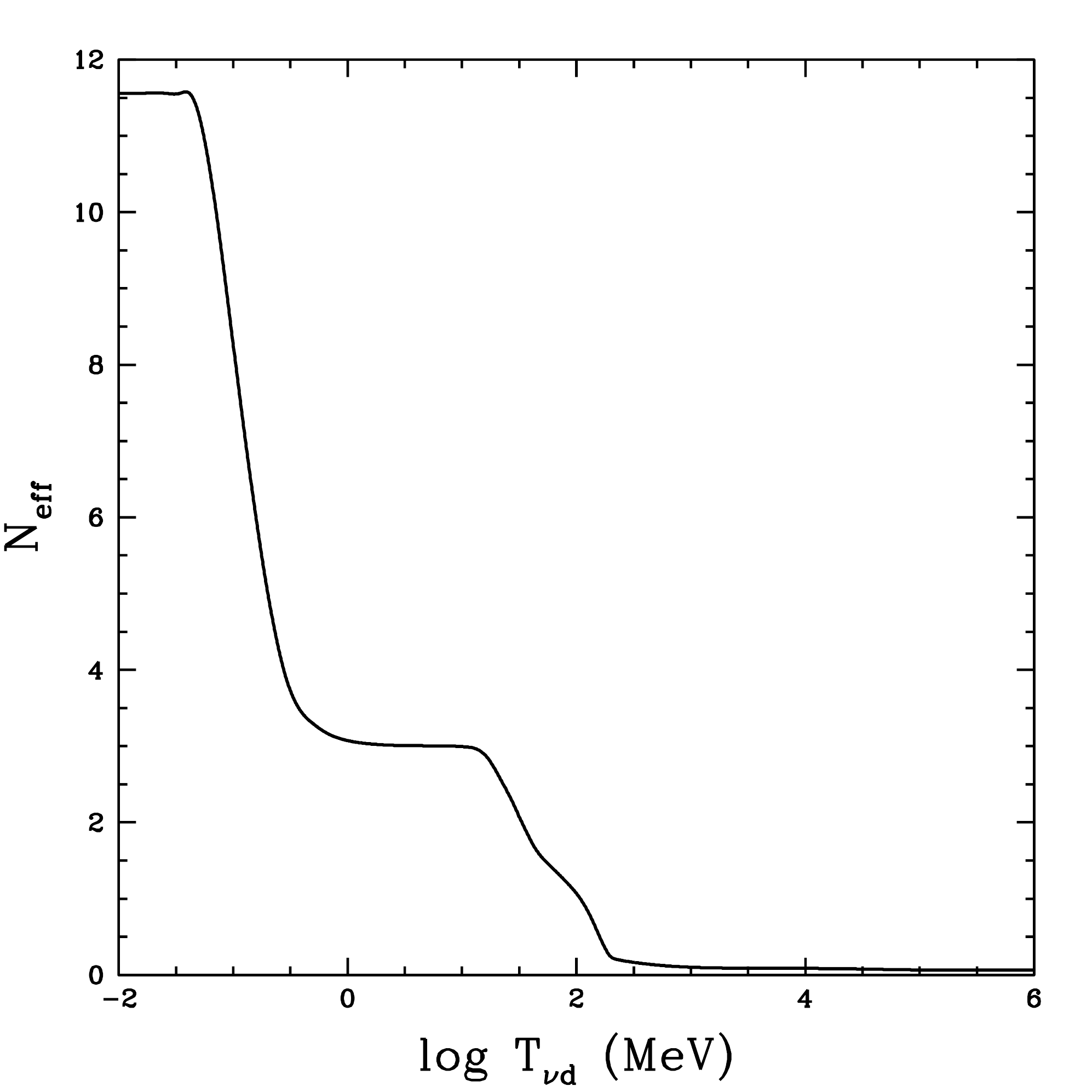}
\caption{The effective number of relativistic degrees of freedom, N$_{\rm eff}$, as a function of $T_{\nu d}$.}
\label{fig:neffvstnud1}
\end{figure}

\subsection{The Effective Number Of Neutrinos\,: $\neff$}

At late times in the early Universe (\eg, after neutrino decoupling but prior to the epoch of matter -- radiation equality and prior to any of the SM neutrinos becoming non-relativistic), the only relativistic SM particles present are the photons and the three SM neutrinos.  As a result, the total energy density (or, the ``radiation" (R) energy density) is,
\beq
\rho_{\rm R} = \rho_{\gamma} + 3\,\rho_{\nu}\,,
\eeq
where $\rho_{\nu}$ is the contribution from one SM neutrino.  After neutrino decoupling and \epm annihilation,
\beq
{\rho_{\nu} \over \rho_{\gamma}} = \bigg({\rho_{\nu} \over \rho_{\gamma}}\bigg)_{0} = {7 \over 8}\bigg({T_{\nu} \over T_{\gamma}}\bigg)^{4}_{0}\,,
\eeq
If $\rho_{\nu 0}^{0}$ is defined to be the value of $\rho_{\nu 0}$ {\it assuming} that $(T_{\nu}/T_{\gamma})^{3}_{0} = 4/11$, then
\beq
\bigg({\rho_{\nu} \over \rho_{\nu}^{0}}\bigg)_{0} = \bigg[{11 \over 4}\bigg({T_{\nu} \over T_{\gamma}}\bigg)^{3}_{0}\bigg]^{4/3}\,,
\eeq
and
\beq
\bigg({\rho_{\rm R} \over \rho_{\gamma}}\bigg)_{0} = 1 + 3\,\bigg({7 \over 8}\bigg)\bigg[{11 \over 4}\bigg({T_{\nu} \over T_{\gamma}}\bigg)^{3}_{0}\bigg]^{4/3}\,,
\eeq
or, normalizing the difference between $\rho_{\rm R}$ and $\rho_{\gamma}$ to $\rho_{\nu}^{0}$,
\beq
\bigg({\rho_{\rm R} - \rho_{\gamma} \over \rho_{\nu}^{0}}\bigg)_{0} = 3\bigg[{11 \over 4}\bigg({T_{\nu} \over T_{\gamma}}\bigg)_{0}^{3}\bigg]^{4/3}\,.
\eeq
This can be generalized from the three SM neutrinos to allow for $\neff$ ``effective neutrinos".  The effective number of neutrinos, $\neff$, here a function of the neutrino decoupling temperature, is defined by,
\beq
{\rm N}_{\rm eff}(T_{\nu d}) \equiv \bigg({\rho_{\rm R} - \rho_{\gamma} \over \rho_{\nu}^{0}}\bigg)_{0} = 3\bigg[{11 \over 4}\bigg({T_{\nu} \over T_{\gamma}}\bigg)_{0}^{3}\bigg]^{4/3}\,.
\eeq

For the SM, assuming instantaneous neutrino decoupling and $g_{e}(T_{\nu d}) = 7/4$ (\eg, massless electrons), N$_{\rm eff} = 3$, corresponding to the three SM neutrinos.  Allowing for the fact that the SM neutrinos don't decouple instantaneously, which enables them to share more of the energy released by \epm annihilation, results in a small ($\sim 1.5\,\%$) increase, N$_{\rm eff} = 3 \rightarrow 3.046$ \cite{mangano}.   

\begin{figure}[!t]
\includegraphics[width=0.60\columnwidth]{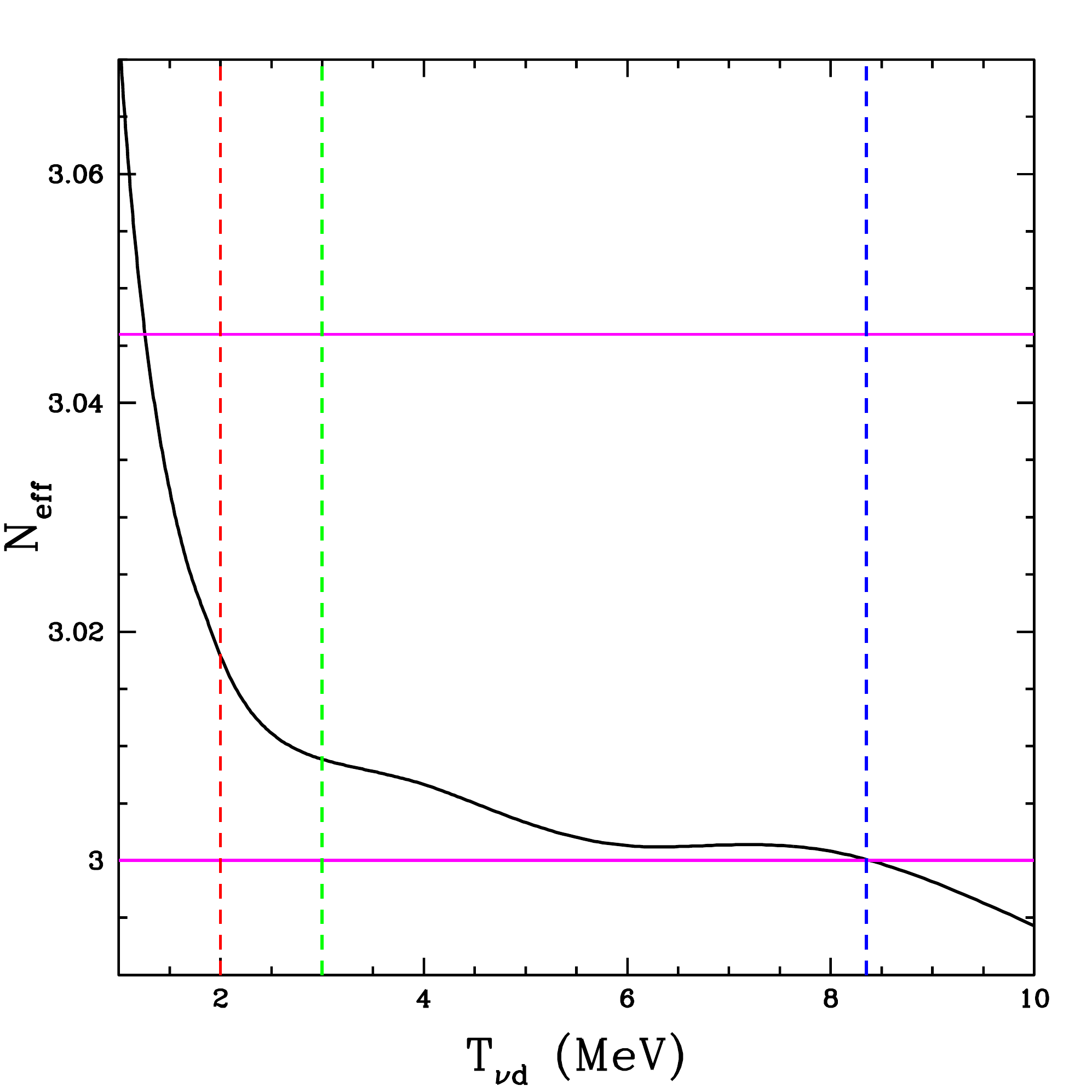}
\caption{A zoomed in version of Fig.\,\ref{fig:neffvstnud1} for $1 \leq T_{\nu d} \leq 10\,{\rm MeV}$ (for a linear temperature scale).  Notice that N$_{\rm eff} = 3$ (lower horizontal, magenta line) when $T_{\nu d} \approx 8.3\,{\rm MeV}$ (dashed, vertical blue line).  For $T_{\nu d} = 2\,{\rm MeV}$, N$_{\rm eff} = 3.018$ (dashed, vertical red line), while for $T_{\nu d} = 3\,{\rm MeV}$, N$_{\rm eff} = 3.012$ (dashed, vertical green line).  In the instantaneous decoupling approximation, N$_{\rm eff} = 3.046$ (upper horizontal, magenta line) when $T_{\nu d} \approx 1.3\,{\rm MeV}$.}
\label{fig:neffvstnud2}
\end{figure}

Since the late time, frozen out ratio of neutrino and photon temperatures depends on the neutrino decoupling temperature, it is informative to allow $T_{\nu d}$ to vary and to explore the dependence of $\neff$ on $T_{\nu d}$.  The relation between $\neff$ and the neutrino decoupling temperature is shown in Fig.\,\ref{fig:neffvstnud1}.  For very high neutrino decoupling temperatures (very weak, weak interactions) the neutrinos are diluted relative to the photons when the latter are heated relative to the decoupled neutrinos by the annihilations and/or decays of all the SM particles.  As a result, for $T_{\nu d} \gg m_{t}$, $g_{s} \rightarrow 427/4$ and N$_{\rm eff} \rightarrow  0.06$.  In the opposite limit of very strong, weak interactions, if the SM neutrinos were to remain coupled through the epoch of \epm annihilation ($T_{\nu d} \ll m_{e}$), sharing the energy released along with the photons, $T_{\nu 0} \rightarrow T_{\gamma 0}$ and N$_{\rm eff} \rightarrow  3(11/4)^{4/3} = 11.56$.

It should be noted that the assumption that $s_{e}/s_{\gamma} = 7/4$ when $T_{\gamma} = T_{\nu d}$, while quite accurate, is not perfect since for all finite temperatures, $s_{e}/s_{\gamma} < 7/4$. Indeed, $s_{e}/s_{\gamma} \rightarrow 7/4$ only in the limit $m_{e}/T_{\nu d} \rightarrow 0$, and while $m_{e}/T_{\nu d} \approx 0.26$ is small, this ratio is not $\ll 1$.  For $T_{\nu d} = 2\,{\rm MeV}$, $s_{e}/s_{\gamma} = 6.95/4$ and $g_{s}(T_{\nu d}) = 42.9/4 = 10.73$.  Fig.\,\ref{fig:neffvstnud2} is a zoomed version of Fig.\,\ref{fig:neffvstnud1}, showing that for the neutrino decoupling temperature adopted here, $T_{\nu d} = 2\,{\rm MeV}$ \cite{enqvist,dolgov,hannestad}, assuming instantaneous neutrino decoupling, N$_{\rm eff} = 3.018$ (if, instead $T_{\nu d} = 3\,{\rm MeV}$ were adopted, N$_{\rm eff} = 3.012$).  Indeed, as may be seen from Fig.\,\ref{fig:neffvstnud2}, the canonical, textbook value of N$_{\rm eff} = 3$ is actually only achieved (in the instantaneous decoupling approximation) for $T_{\nu d} \approx 8\,{\rm MeV}$.  Although this correction ($g_{s}(T_{\nu d}) < 10.75$, N$_{\rm eff} = 3.018$) is small, it is comparable to (within $\sim 40\,\%$ of) the detailed corrections \cite{mangano} accounting, mainly, for non-instantaneous neutrino decoupling.  Indeed, as Fig.\,\ref{fig:neffvstnud1} shows, the longer the neutrinos remain coupled (the stronger the weak interaction), the more they are heated when the \epm pairs annihilate, and the larger is the resulting value of $\neff$.

\subsection{Neutrino Decoupling And the Neutrino Mass Constraint}
\label{numass1}

\begin{figure}[!t]
\includegraphics[width=0.60\columnwidth]{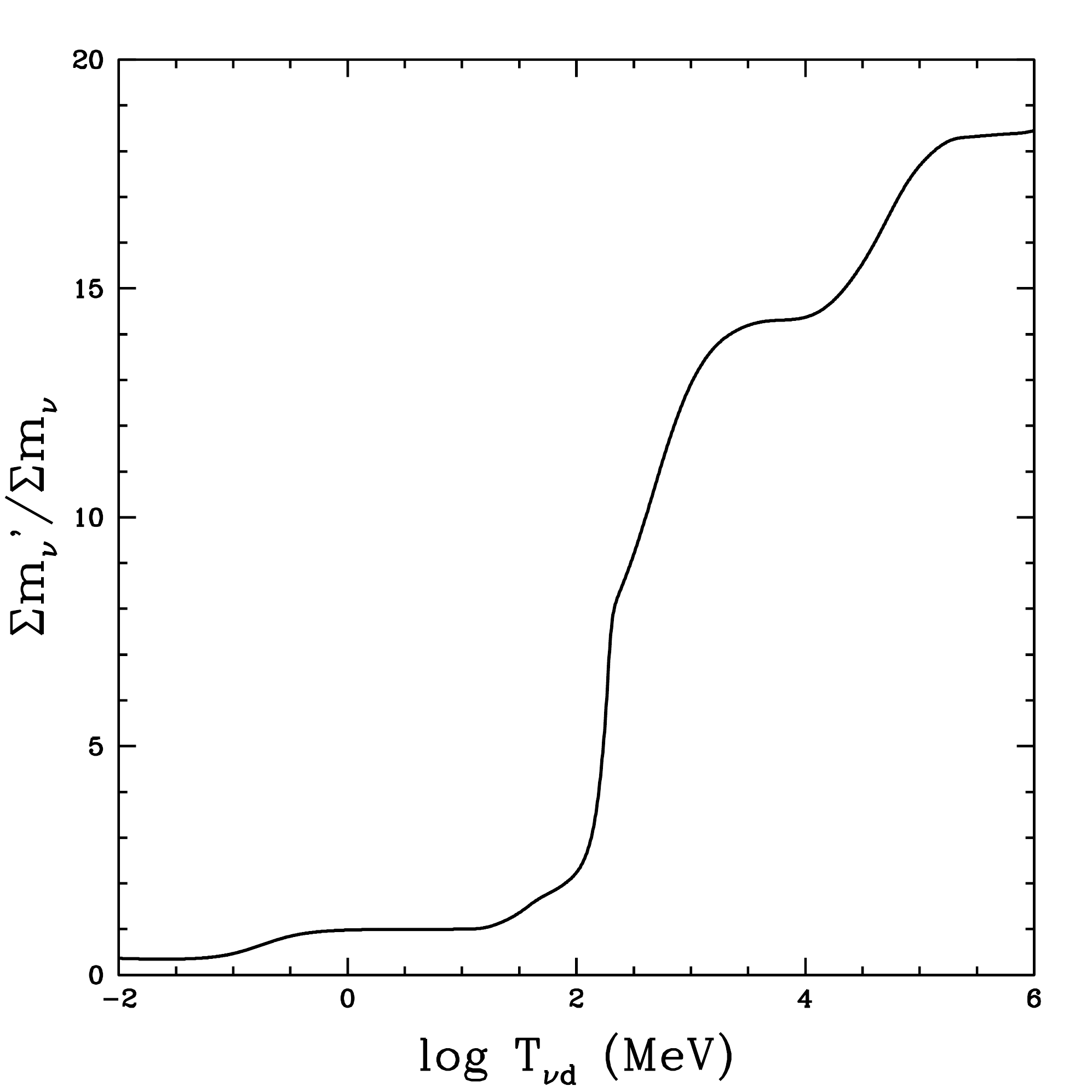}
\caption{The ratio of the sum of the neutrino masses to its canonical value (assuming instantaneous neutrino decoupling and $(T_{\nu}/T_{\gamma})^{3}_{0} = 4/11$), $\Sigma m_{\nu}' /\Sigma m_{\nu}$, as a function of the neutrino decoupling temperature, $T_{\nu d}$.  If the upper bound to $\Sigma\,m_{\nu}$ were $1\,{\rm eV}$, the curve would show the upper bound to the sum of the SM neutrino masses ($\Sigma\,m_{\nu}'$), in eV.}
\label{fig:mnuvstnud1}
\end{figure}

Allowing the neutrino decoupling temperature to vary also has consequences for the CBM/LSS constraint on the sum of the neutrino masses.  Since at least two of the three SM neutrinos have non-zero masses which are large enough so they are non-relativistic in the present Universe, the neutrino contribution to the present Universe mass density is $\rho_{\nu 0} = \Sigma\,m_{\nu}\,n_{\nu 0}$, where $\Sigma\,m_{\nu}$ is the sum of the three SM neutrino masses and $n_{\nu 0}$ is the present number density of one species of the SM neutrinos ($n_{\nu 0} \propto T_{\nu 0}^{3}$).  In the present Universe, the ``frozen out" ratio of the neutrino (each flavor) and photon number densities is
\beq
\bigg({n_{\nu} \over n_{\gamma}}\bigg)_{0} = {3 \over 4}\bigg({T_{\nu} \over T_{\gamma}}\bigg)^{3}_{0} = {3 \over 11}\bigg[{11 \over 4}\bigg({T_{\nu} \over T_{\gamma}}\bigg)^{3}_{0}\bigg]\,.
\eeq
The canonical, instantaneous decoupling result for the SM neutrinos assumes that $(T_{\nu}/T_{\gamma})^{3}_{0} = 4/11$ which, when combined with the number density of CMB photons ($n_{\gamma 0}$) and the critical mass density, leads to a relation between the sum of the SM neutrino masses and $\Omega_{\nu}h^{2}$, $\Sigma\,m_{\nu} = 94.12\,\Omega_{\nu}h^{2}\,{\rm eV}$.  While the detailed calculation of Mangano \etal\,\cite{mangano}, relaxing the instantaneous decoupling approximation, modifies this result to $\Sigma\,m_{\nu} = 93.14\,\Omega_{\nu}h^{2}\,{\rm eV}$, for consistency with the instantaneous decoupling analysis here, this small difference will be ignored.  Here, as $T_{\nu d}$ varies from $\gg 2\,{\rm MeV}$ to $\ll 2\,{\rm MeV}$, $(T_{\nu}/T_{\gamma})^{3}_{0}$ increases from $\ll 4/11$ to 1 ($(n_{\nu}/n_{\gamma})_{0}$ increases from $\ll 3/11$ to $3/4$), modifying the constraint on the sum of the neutrino masses.  If $\Sigma m_{\nu}'$ is defined to be the sum of the neutrino masses when $T_{\nu d}$ is allowed to vary and $\Sigma m_{\nu}$ is the SM quantity, assuming $(n_{\nu}/n_{\gamma})_{0} = 3/11$, then their ratio is a function of the neutrino decoupling temperature,
\beq
{\Sigma m_{\nu}' \over \Sigma m_{\nu}} = {3 \over 11}\bigg({n_{\gamma} \over n_{\nu}}\bigg)_{0} = {4 \over 11}\bigg({T_{\gamma} \over T_{\nu}}\bigg)^{3}_{0} = {2\,g_{s}(T_{\nu d}) - 10.5 \over 11}\,.
\eeq
Fig.\,\ref{fig:mnuvstnud1}, shows $\Sigma m_{\nu}' /\Sigma m_{\nu}$ as a function of $T_{\nu d}$.  For example, if the current CMB and LSS upper bound to the sum of the neutrino masses were $\Sigma m_{\nu} \leq 1\,{\rm eV}$, then the vertical axis in Fig.\,\ref{fig:mnuvstnud1} would be the upper bound to $\Sigma m_{\nu}'$ in eV.  As $T_{\nu d}$ increases from $\ll m_{e}$ to $\gg m_{t}$, $\Sigma m_{\nu}' /\Sigma m_{\nu}$ increases by a factor of $\sim 50$, from $\sim 0.36$ to $\sim 18.5$.

\section{Equivalent Neutrinos}
\label{equivnus}

With the discussion in \S\,\ref{sm} as prelude, the analysis in this section allows for the presence of particles in addition to those provided by the content of the SM.  Along with the SM particles, consider \Deln~additional particles, ``equivalent neutrinos" $\xi$, chosen to be very light ($\Sigma\,m_{\xi} \lsim 10\,{\rm eV}$), or massless, Majorana fermions.  The assumption of a Majorana fermion is for simplicity so that aside from the strength of its coupling to the SM particles, $\xi$ is just like a SM neutrino.  It is important to note that \Deln, a measure of the number of ``extra" neutrinos, is not restricted to be an integer.  In general, \Deln~has discrete values that depend on the nature of the equivalent neutrino and on how many of them are being considered.  For fermionic equivalent neutrinos \Deln~is an integer, while for bosons \Deln~is an integer multiple of 4/7.  For example, \Deln~= 2 for two sterile (Majorana) neutrinos or one Dirac neutrino, while \Deln~= 3 for three right-handed neutrinos, and \Deln~= 4/7 for a scalar.  In the context of the discussion here, ``sterile neutrino" is the special case of an equivalent neutrino that decouples along with the SM neutrinos ($T_{\xi d} = T_{\nu d}$).  The restriction to very light particles is to ensure that the equivalent neutrinos are extremely relativistic when they decouple ($T_{\xi d} \gg m_{\xi}$).

In contrast to the analysis in \S\,\ref{sm}, here the SM neutrino decoupling temperature is fixed at $T_{\nu d} = 2\,{\rm MeV}$, chosen for consistency with most analyses in the literature\,\cite{enqvist,dolgov,hannestad}.  This choice can be modified straightforwardly, \eg, $T_{\nu d} = 3\,{\rm MeV}$, or even for a choice of one decoupling temperature for $\nu_{e}$ (\eg, $T_{\nu ed} = 2\,{\rm MeV}$) and a different one for $\nu_{\mu}$ and $\nu_{\tau}$ (\eg, $T_{\nu \mu d} = T_{\nu \tau d} = 3\,{\rm MeV}$).  The quantitative results for all three choices are very nearly the same.  In contrast to the analysis in \S\,\ref{sm} where $T_{\nu d}$ was allowed to vary, here the free parameter is the equivalent neutrino decoupling temperature, $T_{\xi d}$, assumed to be the same for all (if there are more than one) equivalent neutrinos.  If the equivalent neutrinos are more weakly coupled than are the SM neutrinos they decouple earlier ($T_{\xi d} > T_{\nu d}$), when $g_{s}(T_{\xi d}) > g_{s}(T_{\nu d})$, sharing less of the heating of the SM neutrinos, resulting in $(T_{\xi}/T_{\nu})_{0} < 1$.  On the other hand, if the equivalent neutrinos are more strongly coupled than the SM neutrinos so that $T_{\xi d} < T_{\nu d}$, they remain in equilibrium with the photons and other SM particles to later times, in particular sharing more of the energy/entropy released by the annihilation of the \epm pairs.  This leads to $(T_{\xi}/T_{\nu})_{0} > 1$, along with an increase in $(T_{\nu}/T_{\gamma})_{0}$ from its SM value since the photons now have to share the \epm annihilation energy with the equivalent neutrinos and are cooler than they would be in the absence of the more strongly coupled equivalent neutrinos.  In this case both 
$(T_{\xi}/T_{\gamma})^{3}_{0}$ and $(T_{\nu}/T_{\gamma})^{3}_{0} > 4/11$, so that N$_{eff,\nu} > 3$ and N$_{eff,\xi} > \Delta{\rm N}_{\nu}$, resulting in $\neff > 3 + \Delta{\rm N}_{\nu}$.  

At late times in the early Universe, after the \epm pairs have annihilated, the only particles contributing to the radiation energy density are the photons, the SM neutrinos, and any equivalent neutrinos.  At these times, for $T_{\gamma} \rightarrow T_{\gamma 0} \ll m_{e}$, the radiation energy density, normalized to the energy density in photons alone is,
\beq
\bigg({\rho_{\rm R} \over \rho_{\gamma}}\bigg)_{0} = 1 + {7 \over 8}\bigg[3\bigg({T_{\nu} \over T_{\gamma}}\bigg)^{4}_{0} + \Delta{\rm N}_{\nu}\bigg({T_{\xi} \over T_{\gamma}}\bigg)^{4}_{0}\bigg] = 1 + {7 \over 8}\bigg({T_{\nu} \over T_{\gamma}}\bigg)^{4}_{0}\bigg[3 + \Delta{\rm N}_{\nu}\bigg({T_{\xi} \over T_{\nu}}\bigg)^{4}_{0}\bigg]\,.
\eeq
Recall that the canonical, textbook result is that $(T_{\nu}/T_{\gamma})^{3}_{0} = 4/11$, so that the above result may be written as,
\beq
\bigg({\rho_{\rm R} \over \rho_{\gamma}}\bigg)_{0} \equiv 1 + {7 \over 8}\bigg({4 \over 11}\bigg)^{4/3}{\rm N}_{\rm eff}\,,
\eeq
where the effective number of neutrinos is now a function of both \Deln~and $T_{\xi d}$,
\beq
{\rm N}_{\rm eff} \equiv \bigg[{11 \over 4}\bigg({T_{\nu} \over T_{\gamma}}\bigg)^{3}_{0}\bigg]^{4/3}\bigg[3 + \Delta{\rm N}_{\nu}\bigg({T_{\xi} \over T_{\nu}}\bigg)^{4}_{0}\bigg] = 3\bigg[{11 \over 4}\bigg({T_{\nu} \over T_{\gamma}}\bigg)^{3}_{0}\bigg]^{4/3}\bigg[1 + {\Delta{\rm N}_{\nu} \over 3}\bigg({T_{\xi} \over T_{\nu}}\bigg)^{4}_{0}\bigg]\,.
\eeq
For the canonical result, if the equivalent neutrinos decouple along with the SM neutrinos (\eg, sterile neutrinos) so that $T_{\xi 0} = T_{\nu 0}$, $\neff = 3 + \Delta{\rm N}_{\nu}$.  However, aside from the very small correction $3 \rightarrow 3.018$, $\neff$ generally depends on the combination $\Delta{\rm N}^{*}_{\nu} \equiv \Delta{\rm N}_{\nu}(T_{\xi}/T_{\nu})^{4}_{0}$, which is a function of the equivalent neutrino decoupling temperature $T_{\xi d}$.  There are two interesting regimes, depending on whether $T_{\xi d} \geq T_{\nu d}$ or $T_{\xi d} \leq T_{\nu d}$.

\subsection{Weaker Than Weak Equivalent Neutrinos\,: $T_{\xi d} \geq T_{\nu d}$}

First consider equivalent neutrinos that are more weakly interacting than the SM neutrinos so they decouple before the SM neutrinos, at $T_{\xi d} \geq T_{\nu d}$.  In this case, in the early Universe before neutrino decoupling, when $T_{\gamma} \geq T_{\nu d}$, $T_{\nu} = T_{\gamma}$, while $T_{\xi} \leq T_{\gamma}$.  Early decoupling of any extra, equivalent neutrinos dilutes their contribution to the total energy density, possibly allowing them to avoid the cosmological constraints on $\neff$ (see, \eg,\,\cite{ags}).  Entropy conservation enables us to find the ratio of the $\xi$ to neutrino (and/or photon) temperatures at neutrino decoupling when $T_{\gamma} = T_{\nu d}$,
\beq
\bigg({T_{\xi} \over T_{\nu}}\bigg)^{3}_{T_{\nu d}} = \bigg({T_{\xi} \over T_{\gamma}}\bigg)^{3}_{T_{\nu d}} = {g_{s}(T_{\nu d}) \over g_{s}(T_{\xi d})}\,.
\eeq
The cube of the equivalent neutrino to SM neutrino temperature ratio at the SM neutrino decoupling decreases from $(T_{\xi}/T_{\nu})^{3}_{T_{\nu d}} = 1$ when $T_{\xi d } = T_{\nu d}$ (\eg, for ``sterile" neutrinos), down to  $(T_{\xi}/T_{\nu})^{3}_{T_{\nu d}} = 0.10$ when $T_{\xi d} \gg m_{t}$, corresponding to $(T_{\nu}/T_{\gamma})_{T_{\nu d}}^{4} \approx 0.05$.

As the Universe continues to expand and cool, for $T_{\gamma} < T_{\nu d}$, the ratio of the equivalent neutrino to SM neutrino temperatures is preserved so that for $T_{\gamma} \rightarrow T_{\gamma 0}$,
\beq
\bigg({T_{\xi} \over T_{\nu}}\bigg)^{3}_{0} = \bigg({T_{\xi} \over T_{\nu}}\bigg)^{3}_{T_{\nu d}} = {10.73 \over g_{s}(T_{\xi d})} \leq 1\,.
\eeq

As a result of \epm annihilation the photons are heated relative to both the decoupled SM neutrinos and the equivalent neutrinos, which decoupled earlier.  In this regime ($T_{\xi d} \geq T_{\nu d}$) where both the SM and equivalent neutrinos are decoupled at \epm annihilation, the heating is exactly the same as described in \S\,\ref{sm} so that,
\beq
\bigg({T_{\nu} \over T_{\gamma}}\bigg)^{3}_{0} = {g_{\gamma} \over g_{s}(T_{\nu d}) - 3g_{\nu}} = {2 \over 10.73 - 5.25} = 0.365 \approx 1.004\bigg({4 \over 11}\bigg)\,.
\eeq                                                                  
As a result,
\beq
\bigg({T_{\xi} \over T_{\gamma}}\bigg)^{3}_{0} = \bigg({T_{\nu} \over T_{\gamma}}\bigg)^{3}_{0}\bigg[{g_{s}(T_{\nu d}) \over g_{s}(T_{\xi d})}\bigg] = 0.365\bigg[{g_{s}(T_{\nu d}) \over g_{s}(T_{\xi d})}\bigg] \approx {3.92 \over g_{s}(T_{\xi d})}\,.
\eeq

For the case considered here, the SM neutrinos supplemented by \Deln~equivalent neutrinos which are more weakly coupled to the SM particles than the SM neutrinos,
\beq
{\rm N}_{\rm eff} = {\rm N}_{eff,\nu} + {\rm N}_{eff,\xi} = 3\bigg[{11 \over 4}\bigg({T_{\nu} \over T_{\gamma}}\bigg)^{3}_{0}\bigg]^{4/3} +  \Delta{\rm N}_{\nu}\bigg[{11 \over 4}\bigg({T_{\xi} \over T_{\gamma}}\bigg)^{3}_{0}\bigg]^{4/3}\,,
\eeq
or,
\beq
{\rm N}_{\rm eff}  = \bigg[{11 \over 4}\bigg({T_{\nu} \over T_{\gamma}}\bigg)^{3}_{0}\bigg]^{4/3}(3 + \Delta{\rm N}^{*}_{\nu}) = 3.018\bigg(1 + {\Delta{\rm N}^{*}_{\nu}\over 3}\bigg)\,,
\eeq
where,
\beq
\Delta{\rm N}^{*}_{\nu}  = \Delta{\rm N}_{\nu}\bigg({T_{\xi} \over T_{\nu}}\bigg)^{4}_{0}  = \Delta{\rm N}_{\nu}\bigg({g_{s}(T_{\nu d}) \over g_{s}(T_{\xi d})}\bigg)^{4/3} = \Delta{\rm N}_{\nu}\bigg({10.73 \over g_{s}(T_{\xi d})}\bigg)^{4/3}\,.
\eeq
For one equivalent neutrino, \eg, a Majorana fermion (\Deln~= 1), as $T_{\xi d}$ decreases from $T_{\xi d} \gsim m_{t} \gg T_{\nu d}$ ($g_{s}(T_{\xi d}) \rightarrow 106.75$) to $T_{\xi d} = T_{\nu d}$ ($g_{s}(T_{\xi d}) \rightarrow g_{s}(T_{\nu d}) = 10.73$), the effective number of neutrinos increases from N$_{\rm eff} = 3.065$ to N$_{\rm eff} = 4.024$ (N$_{\rm eff} = 3.018(4/3)$).  These results can be generalized to other choices for the nature and the number of equivalent neutrinos.  

For the special case of sterile neutrinos that decouple along with the SM neutrinos, the results of the Mangano \etal\, analysis \cite{mangano}, relaxing the assumption of instantaneous decoupling, may be appropriate.  If so, then for one (two) sterile neutrinos (\Deln~= 1 (2)), $\neff \rightarrow 3.046(1 + \Delta{\rm N}_{\nu}/3) = 4.06\,(5.08)$.  Since equivalent neutrinos that decouple before the SM neutrinos will not benefit from the additional heating resulting from relaxing the instantaneous decoupling assumption, $\neff = 3.046 + \Delta{\rm N}_{\nu}$ is perhaps more appropriate for them.  However, it is highly unlikely that such small differences will be tested in the foreseeable future.

Since a CMB/LSS constraint on or measurement of N$_{\rm eff}$ results in a constraint on $\Delta{\rm N}_{\nu}^{*}$, which is a function of the equivalent neutrino decoupling temperature, for a fixed value of $\neff$ there is a degeneracy between \Deln~and $T_{\xi d}$.  As an example of this additional freedom, consider the case of three right-handed neutrinos (\Deln~= 3) \cite{ags} which decouple at $T_{\xi d} \approx 180\,{\rm MeV}$, when $g_{s}(T_{\xi d}) \approx 29.6$.  This corresponds to N$_{\rm eff} = 3.80$, consistent with the WMAP 9 year plus SPT results supplemented by information from LSS (\eg, BAO) and measurements of H$_{0}$ \cite{hinshaw,calabrese,hou}.  In contrast, if the same three equivalent neutrinos were to decouple much earlier at $T_{\xi d} \approx 1.5\,{\rm GeV}$, when $g_{s}(T_{\xi d}) \approx 79.3$, this would correspond to N$_{\rm eff} = 3.23$, in excellent agreement with the WMAP 9 year plus ACT results \cite{melchiorri,calabrese}.  A measurement of $\neff < 4$ is not evidence for the absence of one, or even more, equivalent neutrinos.

\subsection{Stronger Than Weak Equivalent Neutrinos\,: $T_{\xi d} < T_{\nu d}$}

While it would seem difficult to have hidden from experimental scrutiny equivalent neutrinos which are more strongly coupled to the SM particles than are the SM neutrinos, for completeness this possibility is explored here.  For more strongly coupled equivalent neutrinos, as $T_{\xi d}$ decreases below $T_{\nu d}$, the equivalent neutrino shares along with the photons some of the energy/entropy released by the \epm annihilations.  However, the decoupled SM neutrinos which have already ``frozen out" prior to $\xi$ decoupling are unheated.  In this regime, when $T_{\gamma} = T_{\nu d}$, $T_{\gamma} = T_{\xi} = T_{\nu}$, while for photon temperatures in the range, $T_{\xi d} \leq T_{\gamma} < T_{\nu d}$, $T_{\gamma} = T_{\xi} \geq T_{\nu}$.  As the temperature decreases further, from $T_{\gamma} = T_{\xi} = T_{\nu d}$ until $\xi$ decoupling when $T_{\gamma} = T_{\xi d}$, the photons and equivalent neutrinos are heated relative to the decoupled neutrinos.  Entropy conservation permits the evaluation of the ratio of neutrino and photon and neutrino and equivalent neutrino temperatures when the equivalent neutrino finally decouples,
\beq
\bigg({T_{\nu} \over T_{\gamma}}\bigg)^{3}_{T_{\xi d}} = \bigg({T_{\nu} \over T_{\xi}}\bigg)^{3}_{T_{\xi d}} = {2g_{s}(T_{\xi  d}) - 7 \over 2g_{s}(T_{\nu d}) - 7} = {2g_{s}(T_{\xi  d}) - 7 \over 14.45}\,,
\eeq
or, in terms of the normalized entropy density in the \epm pairs, $\phi_{e}(x) \equiv s_{e}(x)/s_{e}(0)$, where $x \equiv m_{e}/T$ and $x_{\xi d} \equiv m_{e}/T_{\xi d}$,
\beq
\bigg({T_{\nu} \over T_{\gamma}}\bigg)^{3}_{T_{\xi d}} = {15 + 14\phi_{e}(x_{\xi d}) \over 15 + 14\phi_{e}(x_{\nu d})} = {15 + 14\phi_{e}(x_{\xi d}) \over 28.90}\,.
\eeq
Notice that as $x$ increases from $x \ll 1$ (extremely relativistic) to $x \gg$1 (extremely non-relativistic), $\phi_{e}$ decreases from $1$ to $0$.  Since for the considerations here $T_{\xi d}$ decreases from $2\,{\rm MeV}$ to $\ll m_{e}$, in this regime $\phi_{e}$ decreases from $0.993$ to $0$ and $(T_{\nu}/T_{\gamma})^{3}_{T_{\xi d}}$ decreases from $1$ to $15/28.90 = 0.519$.  In this regime, the photons are less heated than when $T_{\xi d} \geq T_{\nu d}$.

As the Universe continues to expand and cool after the $\xi$ have decoupled ($T_{\gamma} < T_{\xi d}$), the annihilation of any remaining \epm pairs heats the photons relative to the decoupled neutrinos and the now decoupled equivalent neutrinos ($T_{\gamma} \geq T_{\xi} \geq T_{\nu}$) whose temperature ratio remains fixed (\ie, $(T_{\xi}/T_{\nu})_{0} = (T_{\xi}/T_{\nu})_{T_{\xi d}}$).  Entropy conservation in this regime then predicts the frozen out ($T_{\gamma} \rightarrow T_{\gamma 0} \ll m_{e}$) ratio of the equivalent neutrino and photon temperatures,
\beq
\bigg({T_{\xi} \over T_{\gamma}}\bigg)^{3}_{0} = {4 \over 4 + 7\,\phi_{e}(x_{\xi d})}\,.
\eeq
For $T_{\xi d} = T_{\nu d}$, $(T_{\xi}/T_{\gamma})^{3}_{0} = 0.365$.  As $T_{\xi d}$ decreases below $m_{e}$, $\phi_{e} \rightarrow 0$ so that $(T_{\xi}/T_{\gamma})^{3}_{0} \rightarrow 1$ (the equivalent neutrino shares along with the photons all the energy/entropy released by \epm annihilation).
Since the SM neutrinos have already frozen out,
\beq
\bigg({T_{\nu} \over T_{\gamma}}\bigg)^{3}_{0} = \bigg({T_{\nu} \over T_{\xi}}\bigg)^{3}_{T_{\xi d}}\bigg({T_{\xi} \over T_{\gamma}}\bigg)^{3}_{0} = \bigg[{15 + 14\phi_{e}(x_{\xi d}) \over 15 + 14\phi_{e}(x_{\nu d})}\bigg]\bigg[{4 \over 4 + 7\,\phi_{e}(x_{\xi d})}\bigg]\,.
\eeq
As already noted, in this case ($T_{\xi d} < T_{\nu d}$) the SM neutrinos are warmer relative to the photons, than for equivalent neutrinos which decouple before the SM neutrinos because now the photons have to share the \epm energy/entropy with the equivalent neutrinos.  For $T_{\xi d} = T_{\nu d} = 2\,{\rm MeV}$, $(T_{\xi}/T_{\gamma})^{3}_{0} = (T_{\nu}/T_{\gamma})^{3}_{0} = 0.365$.  In contrast, in the limit that $T_{\xi d} \ll m_{e}$, $(T_{\nu}/T_{\gamma})^{3}_{0} \rightarrow 15/28.90 = 0.519$, while $(T_{\xi}/T_{\gamma})^{3}_{0} \rightarrow 1$.

\begin{figure}[!t]
\includegraphics[width=0.60\columnwidth]{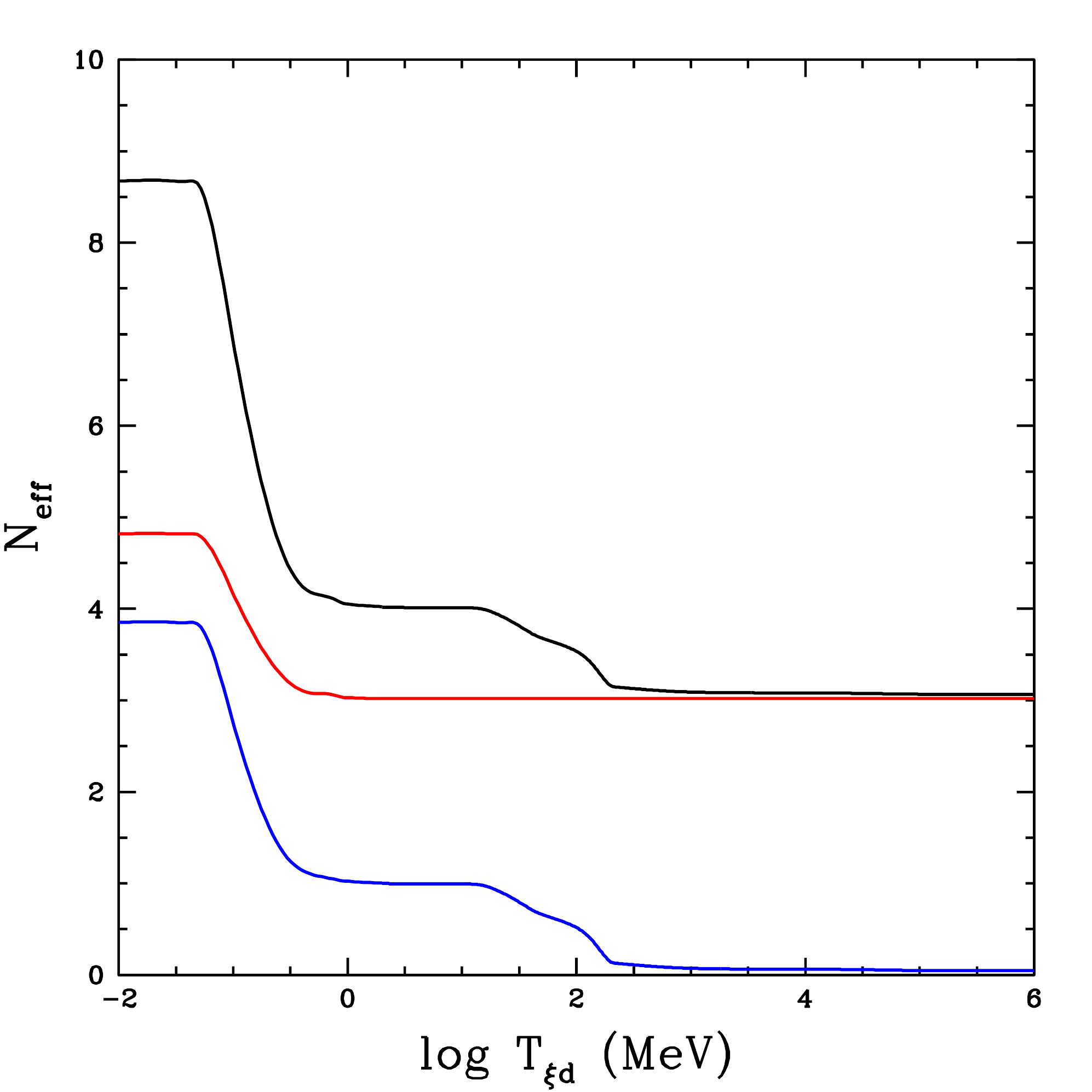}
\caption{Analogous to Fig.\,\ref{fig:neffvstnud1}, N$_{\rm eff}$ is shown as a function of the equivalent neutrino decoupling temperature, $T_{\xi d}$, for one equivalent neutrino, \Deln~= 1 (black curve).  The blue curve is the contribution to N$_{\rm eff}$ from the equivalent neutrino and the red curve is the contribution to N$_{\rm eff}$ from the three SM neutrinos\,.}
\label{fig:neffvstd}
\end{figure}

\begin{figure}[!t]
\includegraphics[width=0.60\columnwidth]{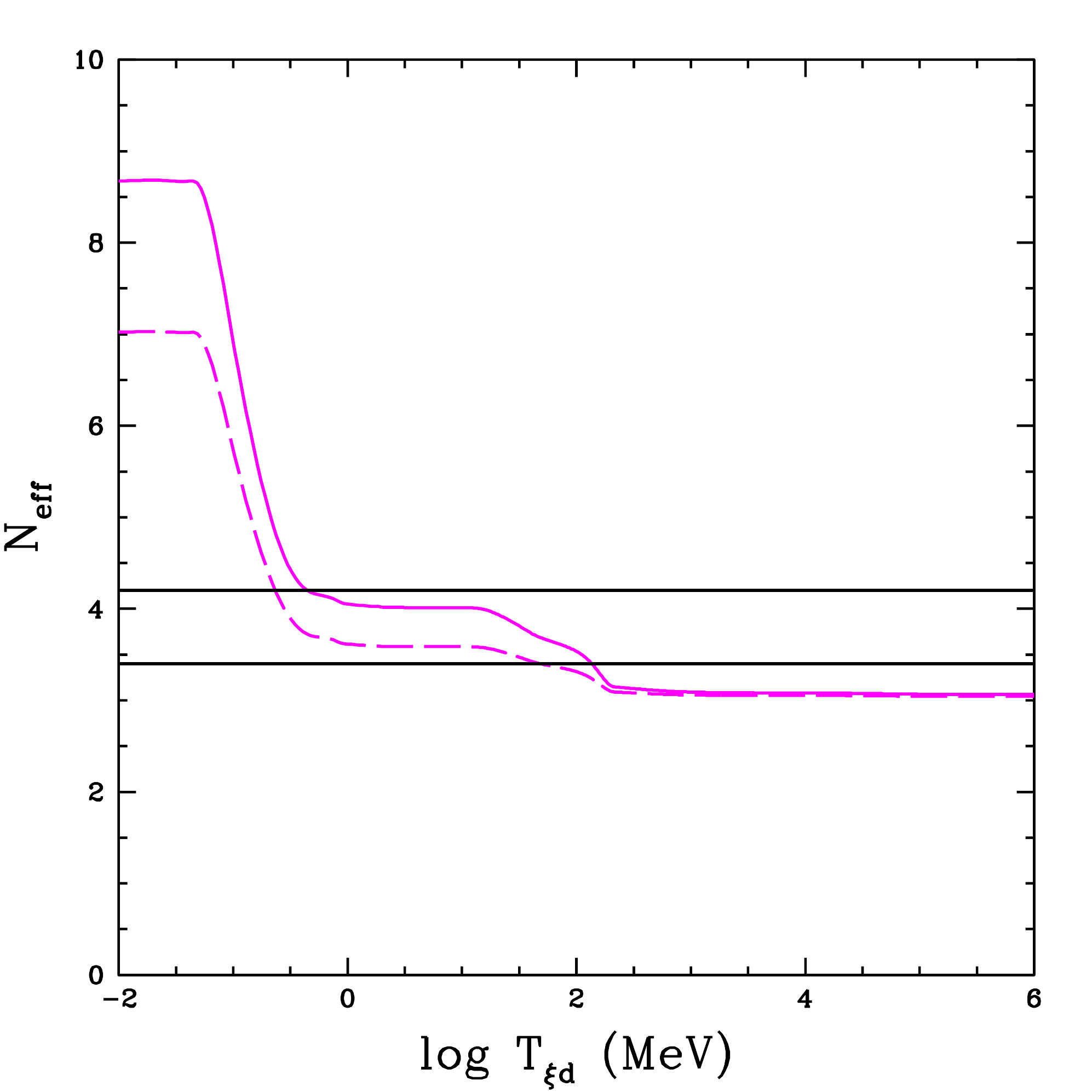}
\caption{N$_{\rm eff}$ is shown as a function of the equivalent neutrino decoupling temperature, $T_{\xi d}$, for one equivalent neutrino, \Deln~= 1, a Majorana fermion (solid curve; the black curve in Fig.\,\ref{fig:neffvstd}).  The long-dashed curve shows N$_{\rm eff}$ for a scalar equivalent neutrino, \Deln~= 4/7.  The horizontal band is the $\pm 1\,\sigma$ region allowed by WMAP9 \cite{hinshaw}.}
\label{fig:neffvstdc}
\end{figure}

As before when $T_{\xi d} > T_{\nu d}$, there are two contributions to N$_{\rm eff}$, from the SM neutrinos (N$_{\rm eff,\nu}$) and from the  \Deln~equivalent neutrinos (N$_{\rm eff,\xi}$),
\begin{multline}
{\rm N}_{\rm eff} = 3\bigg[{11 \over 4}\bigg({T_{\nu} \over T_{\gamma}}\bigg)^{3}_{0}\bigg]^{4/3} + \Delta{\rm N}_{\nu}\bigg[{11 \over 4}\bigg({T_{\xi} \over T_{\gamma}}\bigg)^{3}_{0}\bigg]^{4/3} = \\
3\bigg({11 \over 4 + 7\,\phi_{e}(x_{\xi d})}\bigg)^{4/3}\bigg[\bigg({15 + 14\phi_{e}(x_{\xi d}) \over 15 + 14\phi_{e}(x_{\nu d})}\bigg)^{4/3} + {\Delta{\rm N}_{\nu} \over 3}\bigg]\,.
\end{multline}
In the limit where $T_{\xi d} = T_{\nu d}$ (\eg, for sterile neutrinos), N$_{\rm eff} = 3.018(1 + \Delta{\rm N}_{\nu}/3)$, while in the limit of strongly coupled equivalent neutrinos ($T_{\xi d} \ll m_{e}$), N$_{\rm eff,\nu} = 3\times((11/4)(0.519))^{4/3} = 4.82$ and N$_{\rm eff,\xi} = (11/4)^{4/3}\Delta{\rm N}_{\nu} = 3.85\,\Delta{\rm N}_{\nu}$, so that $\neff = 4.82 + 3.85\,\Delta{\rm N}_{\nu}$\,; for \Deln~= 1, N$_{\rm eff} = 8.67$.  The results for N$_{\rm eff}$ as a function of $T_{\xi d}$ for \Deln~= 1 are shown in Fig.\,\ref{fig:neffvstd}, where the contributions to N$_{\rm eff}$ from the SM neutrinos and the equivalent neutrino are shown separately.  For \Deln~= 1, as the equivalent neutrino decoupling temperature decreases from $T_{\xi d} \gg m_{t}$ to $T_{\xi d} =T_{\nu d}$, N$_{\rm eff}$ increases from 3.07 to 4.02.  As the equivalent neutrino decoupling temperature decreases further, from $T_{\xi d} = T_{\nu d}$ to $T_{\xi d} \ll m_{e}$, N$_{\rm eff}$ increases to 8.67, even though \Deln~= 1.  A measurement of $\neff > 4$ could be consistent with the presence of only one equivalent neutrino (\Deln~= 1).  Note that for a scalar equivalent neutrino, \Deln~= 4/7.  As may be seen in Fig.\,\ref{fig:neffvstdc}, in this case in the limit $T_{\xi d} \ll m_{e}$, $\neff \rightarrow 7.02$.

\subsection{Equivalent Neutrinos And The Neutrino Mass Constraint}
\label{numass2}

\begin{figure}[!t]
\includegraphics[width=0.60\columnwidth]{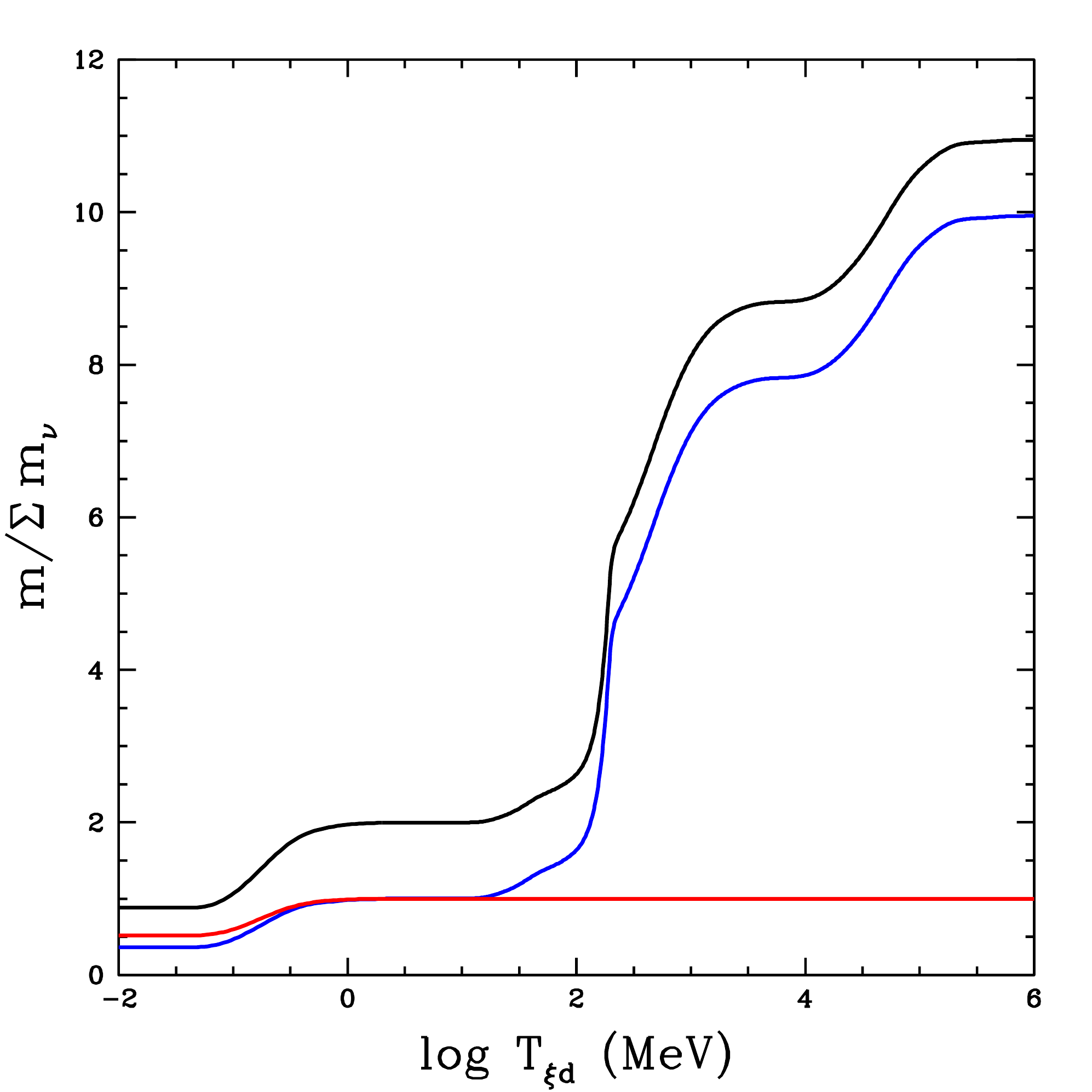}
\caption{Analogous to Fig.\,\ref{fig:mnuvstnud1}, $m \equiv \Sigma\,m_{\xi} + \Sigma\,m_{\nu}'$, normalized to $\Sigma\,m_{\nu}$ ($\Sigma\,m_{\nu} \equiv 94.12\,\Omega_{\nu}h^{2}\,{\rm eV}$), is shown as a function of the equivalent neutrino decoupling temperature, $T_{\xi d}$ (black curve).  The blue curve is for $\Sigma\,m_{\xi}/\Sigma\,m_{\nu}$ and the red curve is for $\Sigma\,m_{\nu}'/\Sigma\,m_{\nu}$.  If the upper bound to $\Sigma\,m_{\nu}$ were $1\,{\rm eV}$, the curves would be the upper bounds to the sums of SM neutrino masses ($\Sigma\,m_{\nu}'$), the equivalent neutrino masses ($\Sigma\,m_{\xi}$), and their sum ($m$), in eV.}
\label{fig:mvstd}
\end{figure}

After \epm annihilation is complete, the present day ratio of the number densities of the SM neutrinos and the equivalent neutrinos to that of the CMB photons is fixed.  For each SM neutrino flavor and for each equivalent neutrino (assuming Majorana fermions),
\beq
\bigg({n_{\nu} \over n_{\gamma}}\bigg)_{0} = {3 \over 4}\bigg({T_{\nu} \over T_{\gamma}}\bigg)^{3}_{0}\, \ \ \,; \ \ \bigg({n_{\xi} \over n_{\gamma}}\bigg)_{0} = {3 \over 4}\bigg({T_{\xi} \over T_{\gamma}}\bigg)^{3}_{0}\,.
\eeq
The present Universe energy density in massive (non-massless) SM neutrinos  and equivalent neutrinos are,
\beq
\rho_{\nu 0} = \Sigma\,m_{\nu}\,n_{\nu 0}\, \ \ {\rm and}\, \ \ \rho_{\xi 0} = \Sigma\,m_{\xi}\,n_{\xi 0}\,.
\eeq 
As before in \S\,\ref{numass1}, if the results for the sum of the neutrino masses ($\Sigma\,m_{\nu}'$) and the sum of the masses of the equivalent neutrinos (if there is more than one, they are assumed to decouple at the same time) $\Sigma\,m_{\xi}$, are compared to those for SM neutrinos which decouple instantaneously at $T_{\nu d} = 2\,{\rm MeV}$ (in which case $\Sigma\,m_{\nu} = 94.12\,\Omega_{\nu}h^{2}\,{\rm eV}$),
\beq
{\Sigma\,m_{\nu}' \over \Sigma\,m_{\nu}} = {4 \over 11}\bigg({T_{\gamma} \over T_{\nu}}\bigg)^{3}_{0}\, \ \ \,, \ \ \ {\Sigma\,m_{\xi} \over \Sigma\,m_{\nu}} = {4 \over 11}\bigg({T_{\gamma} \over T_{\xi}}\bigg)^{3}_{0}\,.
\eeq
Comparing with the results of the previous section, these results may be also written as,
\beq
{\Sigma\,m_{\nu}' \over \Sigma\,m_{\nu}} = \bigg({3 \over N_{eff,\nu}}\bigg)^{3/4}\, \ \ \,, \ \ \ {\Sigma\,m_{\xi} \over \Sigma\,m_{\nu}} = \bigg({1 \over N_{eff,\xi}}\bigg)^{3/4}\,.
\eeq
Since a constraint on the current energy density in hot, dark matter ($\Omega_{\rm HDM}$), leads to a constraint on the sum of the SM and equivalent neutrino masses, in the presence of equivalent neutrinos, this neutrino mass constraint is modified,
\beq
m \equiv \Sigma\,m_{\xi} + \Sigma\,m_{\nu}' \leq 94.12\,\Omega_{\rm HDM}h^{2}\bigg({\Sigma\,m_{\nu}' \over \Sigma\,m_{\nu}} + {\Sigma\,m_{\xi} \over \Sigma\,m_{\nu}}\bigg)\,{\rm eV} = 94.12\,\Omega_{\rm HDM}h^{2}\bigg[\bigg({3 \over N_{eff,\nu}}\bigg)^{3/4} + \bigg({1 \over N_{eff,\xi}}\bigg)^{3/4}\bigg]\,{\rm eV}\,.
\eeq
These results for the SM neutrino and equivalent neutrino masses as well as for their sum are shown in Fig.\,\ref{fig:mvstd} as a function of the equivalent neutrino decoupling temperature.  For example, if observations should find $94.12\,\Omega_{\rm HDM}h^{2} = 1$, corresponding to $\Sigma\,m_{\nu} \leq 1\,{\rm eV}$, then the vertical scale in Fig.\,\ref{fig:mvstd} is the upper bound to the sum of the SM neutrino and equivalent neutrino masses, in eV.  Notice that for very weakly coupled equivalent neutrinos any CMB/LSS constraint on the sum of the neutrino masses is relaxed by $\sim$ an order of magnitude (in this limit, $\Sigma\,m_{\nu}' \approx \Sigma\,m_{\nu}$, while $\Sigma\,m_{\xi} \approx 10\,\Sigma\,m_{\nu}$).  However, for sterile neutrinos ($T_{\xi d} = T_{\nu d} = 2\,{\rm MeV}$), the mass constraint is relaxed by only a factor of two (see Fig.\,\ref{fig:mvstd}).

\section{Light Or Very Light WIMPs\,: Rescuing Sterile Neutrinos}
\label{light wimp}

In this section the effect on $\neff$ of the presence of a WIMP, sufficiently light so that its late time annihilation heats the photons beyond the usual heating from \epm annihilation, is investigated. While the dark matter candidates ($\chi$) supplied by most supersymmetric models tend to be very massive, $m_{\chi} \gsim$ tens or hundreds of GeV, in recent years there has also been interest in the light ($m_{e} \lsim m_{\chi} \lsim$ tens of MeV) or very light ($m_{\chi} \lsim m_{e}$) WIMPs \cite{ktw,serpico,boehm1,boehm2,boehm3,hooper1,boehm4,hooper2,ahn,fayet,hooper3,feng} considered here. The discussion in this section has some overlap with earlier work of Kolb \etal\,\cite{ktw} and of Serpico and Raffelt \cite{serpico}, and with the recent analyses of Ho and Scherrer \cite{hoscherrer1,hoscherrer2}.  Assume, initially, that there are no equivalent neutrinos (\Deln~= 0), but there is a light WIMP, $\chi$, a Majorana fermion (to be generalized later to a WIMP that is a Dirac fermion or a scalar boson, and to $\Delta{\rm N}_{\nu} \neq 0$).  The annihilation of a WIMP more massive than $\sim 20\,{\rm MeV}$ occurs prior to the decoupling of the SM neutrinos, heating them along with the photons and the \epm pairs present at that time, preserving the standard results discussed in \S\,\ref{sm}.  Note that it is essential here to assume that the light WIMP couples to photons and \epm pairs but does not couple to the SM neutrinos since through such coupling the neutrinos could be kept in equilibrium with the photons, leading to $(T_{\nu}/T_{\gamma})_{0} \rightarrow 1$ and N$_{\rm eff} \rightarrow 3\,(11/4)^{4/3} = 11.56$ (see \S\,\ref{sm}).  This assumption will be reversed in the next section where WIMPs that couple only to the SM neutrinos are considered.  In the presence of ``massive" light WIMPs ($m_{\chi} \gsim 20\,{\rm MeV}$) there is no change from the standard result that for \Deln~= 0, N$_{\rm eff} = 3.018\,({\rm or}, 3.046$\,\cite{mangano}).  However, the late time annihilation of sufficiently light WIMPs ($m_{\chi} \lsim 12\,{\rm MeV} \approx 6\,T_{\nu d}$) will further heat the photons relative to the now decoupled neutrinos, resulting in photons that are hotter than the SM neutrinos in the absence of the light WIMP.  This dilutes the contribution of the SM neutrinos to the early Universe energy density, leading to the surprising result that, even in the presence of the three SM neutrinos, N$_{\rm eff} < 3$.  This opens the door for \Deln~$> 0$ to be consistent with a measurement of $\neff = 3$. 

\begin{figure}[!t]
\begin{center}
\includegraphics[width=0.60\columnwidth]{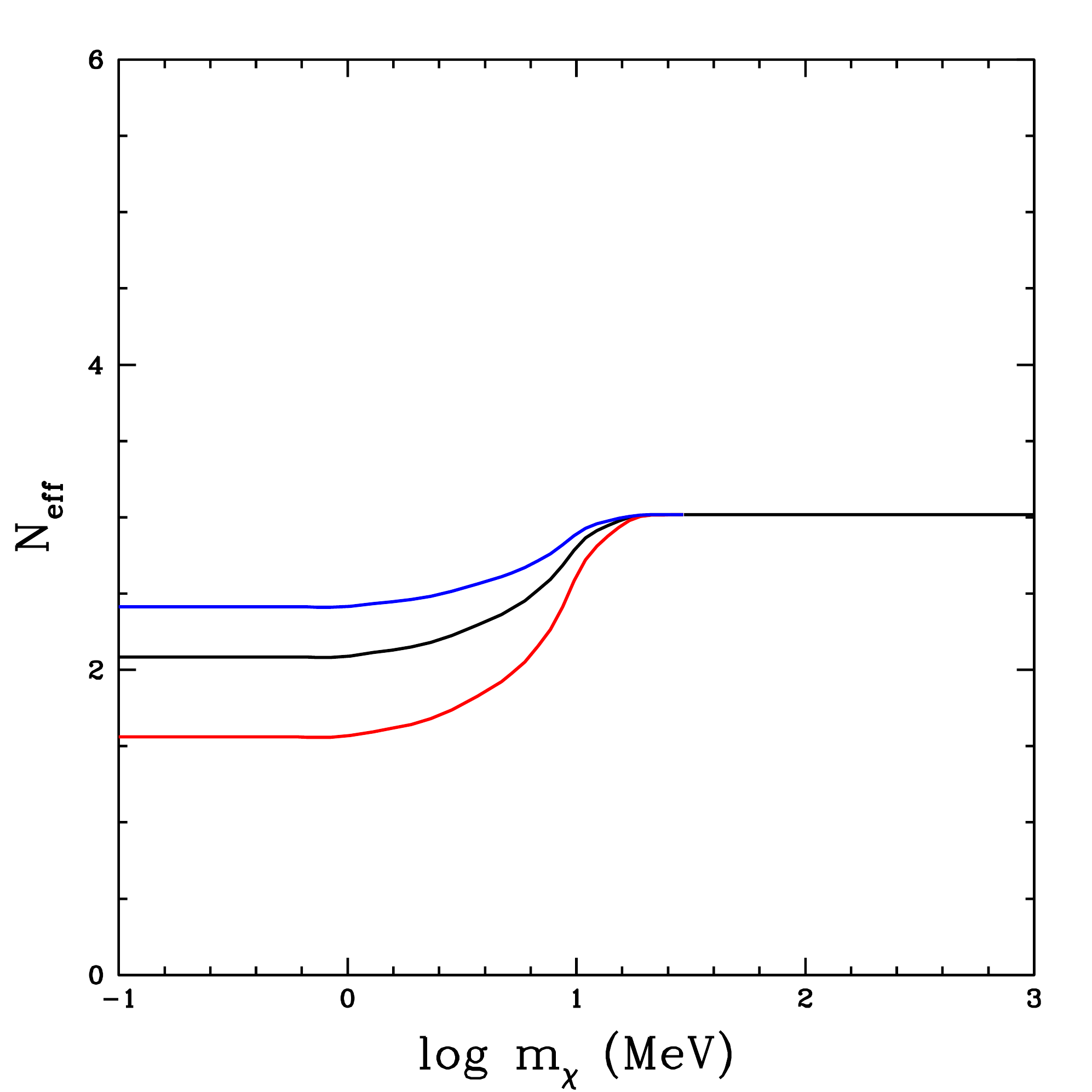}
\caption{The effective number of neutrinos as a function of the WIMP mass for a Majorana WIMP (black), a Dirac WIMP (red), and a scalar WIMP (blue).}
\label{fig:neffvsma}
\end{center}
\end{figure}

As before, the late time ratio of neutrino to photon temperatures, $(T_{\nu}/T_{\gamma})_{0}$, may be evaluated by comparing the entropy in a comoving volume at $T_{\gamma} = T_{\nu d}$ with the same quantity evaluated at $T_{\gamma} = T_{\gamma 0} \ll m_{e}\,(m_{\chi})$.  At late times, 
\beq
\bigg({T_{\nu} \over T_{\gamma}}\bigg)^{3}_{0} = {2 \over 2 + {7 \over 2}\phi_{ed} + {7 \over 4}\phi_{\chi d}}\,,
\eeq
where $\phi(x) \equiv s(x)/s(0)$ (for fermions; the same for Majorana and Dirac fermions) and $\phi_{ed}$ is evaluated at $x_{ed} = m_{e}/T_{\nu d}$ while $\phi_{\chi d}$ is evaluated at $x_{\chi d} = m_{\chi}/T_{\nu d}$.  It is usually assumed that $\phi_{ed} = 1$ but, as seen above in \S\ref{sm}, for $T_{\nu d} = 2\,{\rm MeV}$, $\phi_{ed} = 0.993$.  For consistency, this latter value is adopted here (along with the assumption of instantaneous decoupling) resulting in,
\beq
\bigg({T_{\nu} \over T_{\gamma}}\bigg)^{3}_{0} = {4 \over 10.95 + {7 \over 2}\phi_{\chi d}}\, \ \ \,, \ \ \ {11 \over 4}\bigg({T_{\nu} \over T_{\gamma}}\bigg)^{3}_{0} = {11 \over 10.95 + 3.5\,\phi_{\chi d}}\,.
\label{eq:entropy}
\eeq
As a result,
\beq
{\rm N}_{\rm eff} \equiv 3\bigg[{11 \over 4}\bigg({T_{\nu} \over T_{\gamma}}\bigg)^{3}_{0}\bigg]^{4/3} = 3\bigg[{11 \over 10.95 + 3.5\,\phi_{\chi d}}\bigg]^{4/3} \leq 3.018\,.
\label{eq:neff}
\eeq

\begin{figure}[!t]
\begin{center}
\includegraphics[width=0.60\columnwidth]{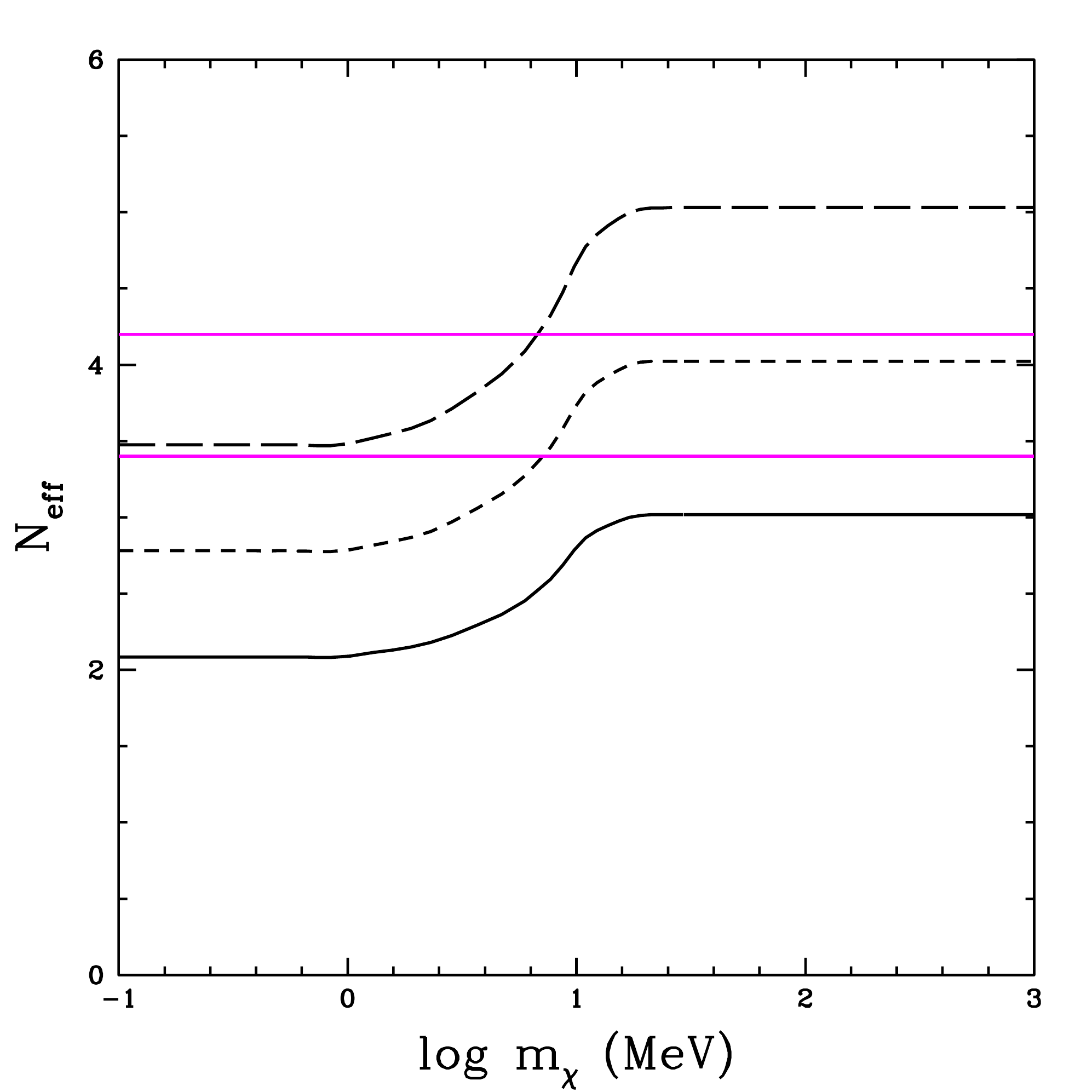}
\caption{The effective number of neutrinos, N$_{\rm eff}$, as a function of the WIMP mass for a Majorana WIMP.  The solid curve is for the case of no sterile neutrinos.  The short-dashed curve is for one sterile neutrino.  The long dashed curve is for two sterile neutrinos.  The horizontal (purple) lines show the $\pm\,1\sigma$ band allowed by the WMAP 9 year data \cite{hinshaw}.}
\label{fig:neffvsm3}
\end{center}
\end{figure}

In the presence of a light WIMP, $\neff$ is a function of the light WIMP mass through the dependence of $\phi_{\chi d}$ on $x_{\chi d} = m_{\chi}/T_{\nu d}$.  In the limit of ``high" light WIMP masses, $m_{\chi} \gg T_{\nu d}$, $\phi_{\chi d} \rightarrow 0$ and, N$_{\rm eff} \rightarrow 3.018$, recovering the SM result.  However, in the opposite limit, for very light WIMPs with $m_{\chi} \ll m_{e} \lsim T_{\nu d}/4$, $\phi_{\chi d} \rightarrow 1$ and, N$_{\rm eff} \rightarrow 2.085$ \cite{hoscherrer1}.  The evolution of N$_{\rm eff}$ with $m_{\chi}$ is shown by the black curve in Fig.\,\ref{fig:neffvsma}.

\begin{figure}[!t]
\begin{center}
\includegraphics[width=0.60\columnwidth]{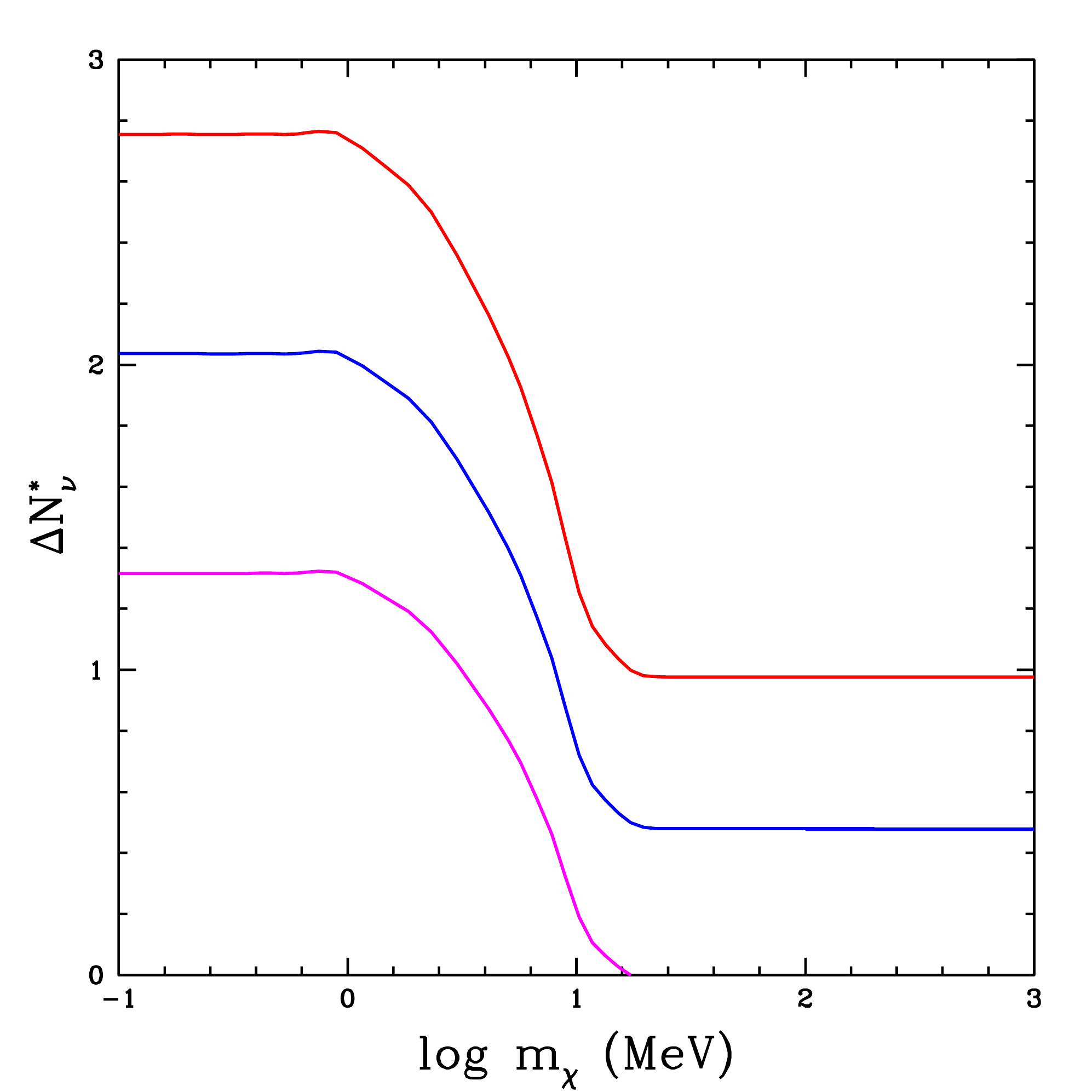}
\caption{The effective number of equivalent neutrinos, $\Delta{\rm N}_{\nu}^{*}$ (see the text), as a function of the WIMP mass, $m_{\chi}$, for a Majorana WIMP, consistent with an observationally determined value of N$_{\rm eff}$.  The purple curve is for N$_{\rm eff} = 3.0$, the blue curve is for N$_{\rm eff} = 3.5$, the red curve is for N$_{\rm eff} = 4.0$.}
\label{fig:delnvsm}
\end{center}
\end{figure}

This result is for a WIMP that is a Majorana fermion.  It is straightforward to generalize this result to a WIMP that is a Dirac fermion, or for bosons \cite{hoscherrer1,hoscherrer2}, by rewriting the entropy conservation equation (Eq.\,\ref{eq:entropy}) as,
\beq
\bigg({T_{\nu} \over T_{\gamma}}\bigg)^{3}_{0} = {2 \over 2 + {7 \over 2}\phi_{ed} + \tilde{g}_{\chi}\phi_{\chi d}}\,,
\eeq
where $\tilde{g}_{\chi} = 7/4$ for a Majorana WIMP, 7/2 for a Dirac WIMP, and 1 for a scalar WIMP; a vector boson WIMP would have $\tilde{g}_{\chi} = 3$.  However, note that for bosons, the quantity $\phi = s(x)/s(0)$, which has been derived for the Majorana and Dirac WIMPs using the Fermi-Dirac distribution, must be replaced with the corresponding function evaluated using the Bose-Einstein distribution.  The results for these different choices are shown in Fig.\,\ref{fig:neffvsma}.  In the limit of ``high" WIMP mass, $m_{\chi} \gsim 12\,{\rm MeV}$, all these cases approach N$_{\rm eff} \approx 3.02$, but they differ for very light WIMPs with $m_{\chi} \lsim 1\,{\rm MeV}$.  While $\neff \rightarrow 2.09$ for a Majorana WIMP, for a Dirac WIMP, N$_{\rm eff} \rightarrow 1.56$, and for a scalar WIMP, N$_{\rm eff} \rightarrow 2.41$.  As may be seen in Fig.\,\ref{fig:neffvsma}, the transition from the ``standard" value of N$_{\rm eff} \approx 3$ in the absence of extra equivalent neutrinos or dark radiation, to the asymptotic values of N$_{\rm eff} < 3$ occurs over a relatively small range in the light WIMP mass, $1 \lsim m_{\chi} \lsim 12\,{\rm MeV}$.  In the absence of ``dark radiation" or equivalent neutrinos (\Deln~= 0), the presence of a sufficiently light WIMP allows the effective number of neutrinos to take on any value from N$_{\rm eff} \approx 1.56$ to N$_{\rm eff} \approx 3.02$, depending on the nature of the WIMP and its mass.

\subsection{Light WIMP And Sterile Neutrinos: Degeneracy Between $m_{\chi}$ And \Deln}

\begin{figure}[!t]
\begin{center}
\includegraphics[width=0.60\columnwidth]{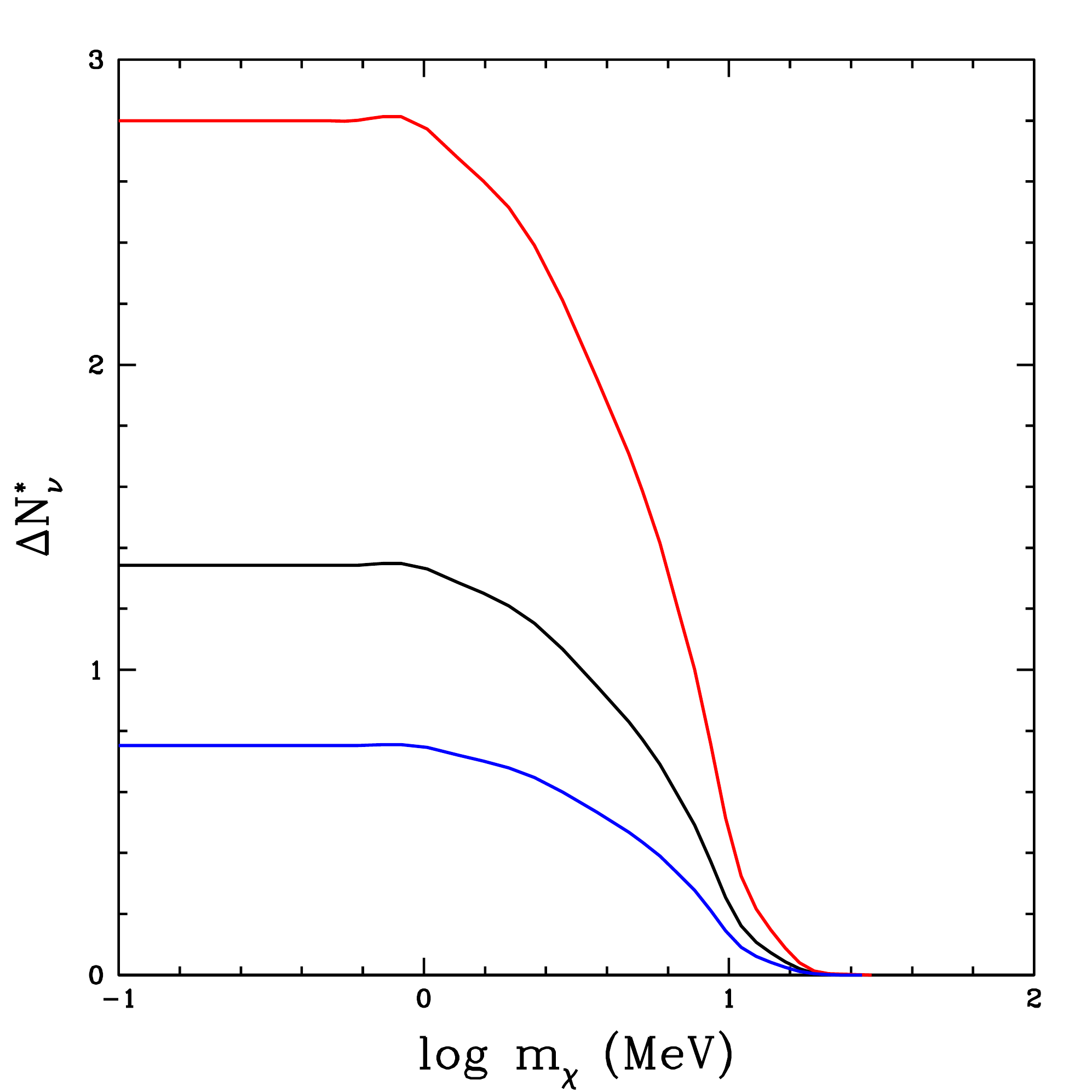}
\caption{The analog of Fig.\,\ref{fig:delnvsm} for the choice of $\neff = 3$.  The black curve is for a Majorana fermion WIMP, the red curve is for a Dirac fermion WIMP, and the blue curve is for a scalar boson WIMP (see Fig.\,\ref{fig:neffvsma}).}
\label{fig:delnvsma}
\end{center}
\end{figure}

To explore how $\neff$ changes in the presence of both a light WIMP ($\chi$) and equivalent neutrinos ($\xi$), allow for \Deln~equivalent neutrinos that, for simplicity, all decouple at the same temperature, $T_{\xi d}$.  If N$_{\rm eff}^{0} \equiv {\rm N}_{\rm eff}(\Delta{\rm N}_{\nu} = 0)$ (see Eq.\,\ref{eq:neff}) and N$_{\rm eff} \equiv {\rm N}_{\rm eff}(\Delta{\rm N}_{\nu} \neq 0)$ then,
\beq
{\rm N}_{\rm eff} = \bigg[3 + \Delta{\rm N}_{\nu}\bigg({T_{\xi} \over T_{\nu}}\bigg)^{4}_{0}\bigg]\bigg[{11 \over 4}\bigg({T_{\nu} \over T_{\gamma}}\bigg)_{0}^{3}\bigg]^{4/3} = {\rm N}_{\rm eff}^{0}\bigg[1 + {\Delta{\rm N}_{\nu} \over 3}\bigg({T_{\xi} \over T_{\nu}}\bigg)^{4}_{0}\bigg] = {\rm N}_{\rm eff}^{0}\bigg(1 + {\Delta{\rm N}^{*}_{\nu} \over 3}\bigg)\,.
\eeq

Suppose there are one or even two sterile neutrinos, so that $(T_{\xi}/T_{\nu})_{0} = 1$ and $\neff = \neff^{0}(1 + \Delta{\rm N}_{\nu}/3)$.  Depending on the nature of the WIMP and its mass it is possible to account for any value of N$_{\rm eff}$ in the range $2.08 \lsim {\rm N}_{\rm eff} \lsim 4.02$ (for one sterile neutrino) or $2.60 \lsim{\rm N}_{\rm eff} \lsim 5.03$ (for two sterile neutrinos).  Since in the presence of sterile neutrinos the effective number of neutrinos depends on the WIMP mass and its nature, along with the number of sterile neutrinos, $\neff = {\rm N}_{\rm eff}(m_{\chi}\,,\Delta{\rm N}_{\nu})$, there is a degeneracy between the number of sterile neutrinos and the WIMP mass (and its nature).  The same observationally determined value of the effective number of neutrinos can be achieved with different combinations of the light WIMP mass and the number of sterile neutrinos.  This degeneracy is illustrated in Fig.\,\ref{fig:neffvsm3} for a Majorana fermion WIMP.  As seen in Fig.\,\ref{fig:neffvsm3}, for one sterile neutrino ($\Delta{\rm N}_{\nu} = 1$), ${\rm N}_{\rm eff} = 4{\rm N}_{\rm eff}^{0}/3$\,; for two sterile neutrinos ($\Delta{\rm N}_{\nu} = 2$), ${\rm N}_{\rm eff} = 5{\rm N}_{\rm eff}^{0}/3$.  For example, as shown in Fig.\,\ref{fig:neffvsm3}, for a sufficiently low mass Majorana fermion WIMP, $m_{\chi} \lsim 1\,{\rm MeV}$, N$_{\rm eff}^{0} \rightarrow 2.09$, so that for one (two) sterile neutrino(s), ${\rm N}_{\rm eff} = 2.78\,(3.48)$, consistent with current CMB/LSS constraints\,\cite{hinshaw,melchiorri,hou,calabrese}.  A CMB/LSS determination of N$_{\rm eff} \approx 3$ does not, by itself, exclude the possibility of one sterile neutrino.  Indeed, the current CMB/LSS data appear to favor one, or possibly two, sterile neutrinos.

\subsection{Light WIMP and Equivalent Neutrinos: Degeneracy Among $m_{\chi}$, \Deln, And $T_{\xi d}$}
\label{degeneracy}

Current pre-Planck constraints on N$_{\rm eff}$ from WMAP9, ACT3, and SPT, supplemented by LSS data from BAO and measurements of $H_{0}$ are consistent with values for the effective number of neutrinos in the range, $3 \lsim {\rm N}_{\rm eff} \lsim 4$ \cite{hinshaw,melchiorri,hou,calabrese}.  Values of N$_{\rm eff}$ in this range can be achieved by different combinations of N$_{\rm eff}^{0}(m_{\chi})$ and $\Delta{\rm N}_{\nu}^{*} \equiv \Delta{\rm N}_{\nu}\,(T_{\xi}/T_{\nu})^{4}_{0}$.  If the restriction to sterile neutrinos is relaxed so that $(T_{\xi}/T_{\nu})_{0} \neq 1$, it is $\Delta{\rm N}_{\nu}^{*}$ and the WIMP mass that are degenerate\,: ${\rm N}_{\rm eff}(m_{\chi}\,,\Delta{\rm N}_{\nu}^{*}) = {\rm N}_{\rm eff}^{0}(m_{\chi})(1 + \Delta{\rm N}_{\nu}^{*}/3)$.  For example, if the equivalent neutrinos decouple prior to the decoupling of the SM neutrinos so that $(T_{\xi}/T_{\nu})_{0} \leq 1$, the contribution of the equivalent neutrinos to $\neff$ is diluted, $\Delta{\rm N}_{\nu}^{*} \leq \Delta{\rm N}_{\nu}$.  $\Delta{\rm N}_{\nu}^{*}$ is shown as a function of the WIMP mass in Fig.\,\ref{fig:delnvsm} for three different choices of N$_{\rm eff} = 3.0,\,3.5,\,4.0$, demonstrating that depending on the WIMP mass, these values of $\neff$ are consistent with $\Delta{\rm N}_{\nu}^{*}$ in the range $0 \leq \Delta{\rm N}_{\nu}^{*} \lsim 2.8$.  As an illustrative example, reconsider the case of three right-handed neutrinos\,\cite{ags} (see \S\,III\,A), so that \Deln~= 3 and $\Delta{\rm N}_{\nu}^{*} = 3(T_{\xi}/T_{\nu})^{4}_{0}$.  For $m_{\chi} \lsim 1\,{\rm MeV}$, $\neff = 3$ requires $\Delta{\rm N}_{\nu}^{*} \approx 1.3$, or $(T_{\xi}/T_{\nu})^{4}_{0} \approx 0.4$, which is achieved for $T_{\xi d} \approx 120\,{\rm MeV} \approx 60\,T_{\nu d}$ \cite{ags}.  A determination of $\neff = 3$ does not exclude three, right-handed neutrinos.  This is illustrated in Fig.\,\ref{fig:delnvsma} where $\neff = 3$ is adopted and $\Delta{\rm N}_{\nu}^{*}$ is shown as a function of the WIMP mass (as in Fig.\,\ref{fig:delnvsm}) for Majorana and Dirac fermion WIMPs as well as for a scalar WIMP.  As may be seen in Fig.\,\ref{fig:delnvsma}, for $m_{\chi} \lsim 1\,{\rm MeV}$, $\Delta{\rm N}_{\nu}^{*} > 0$ even though $\neff = 3$.  The absence of evidence for equivalent neutrinos ($\neff = 3$) is not evidence for the absence of equivalent neutrinos.

\subsection{SM And Sterile Neutrino Masses In The Presence Of A Light WIMP}
\label{numass3}

\begin{figure}[!t]
\begin{center}
\includegraphics[width=0.60\columnwidth]{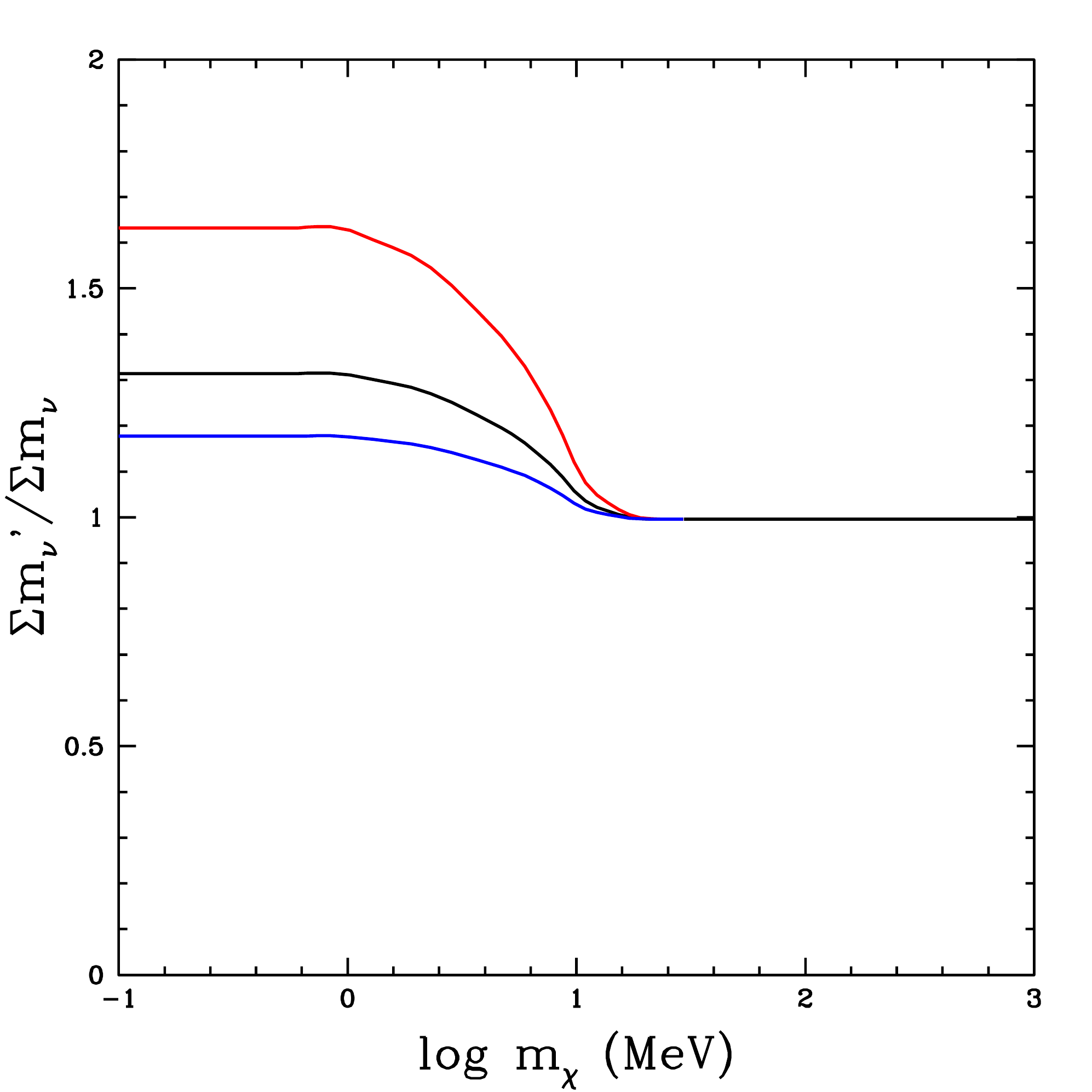}
\caption{The ratio of the sum of the SM plus sterile neutrino masses to its canonical value, $\Sigma m_{\nu}' /\Sigma m_{\nu}$, assuming instantaneous neutrino decoupling and $(T_{\nu}/T_{\gamma})_{0}^{3} = 4/11$, as a function of the WIMP mass for a Majorana WIMP (black), a Dirac WIMP (red), and a scalar WIMP (blue).  If the upper bound to $\Sigma\,m_{\nu}$ were $1\,{\rm eV}$, the curves would be the upper bounds to the sum of the SM plus sterile neutrino masses, in eV.}
\label{fig:mnuvsm}
\end{center}
\end{figure}

As has been noted in \S\,\ref{numass1} and \S\,\ref{numass2} above, the constraint on the sum of the neutrino masses is modified if the late time ($T_{\gamma} \rightarrow T_{\gamma 0}$) ratio of neutrino to photon temperatures changes,\beq
{\Sigma\,m_{\nu}' \over \Sigma\,m_{\nu}} = {4 \over 11}\bigg({T_{\gamma} \over T_{\nu}}\bigg)^{3}_{0} = \bigg({3 \over N_{\rm eff}^{0}}\bigg)^{3/4}\,.
\eeq
Here, for simplicity, it is assumed that the \Deln~extra neutrinos decouple along with the SM neutrinos (\eg, they are sterile neutrinos) so that $\Sigma\,m_{\nu}'$ is the sum of the SM and sterile neutrino masses.  Since for light WIMPs $\neff^{0} \leq 3$, this allows $\Sigma\,m_{\nu}' \geq \Sigma\,m_{\nu}$, permitting more massive SM neutrinos to be compatible with current CMB/LSS constraints.  Compared to the examples discussed earlier (\S\,\ref{numass1} and \S\,\ref{numass2}), for the case considered here of light WIMPs, with or without sterile neutrinos, the deviation of N$_{\rm eff}^{0}$ from 3 is less dramatic, resulting in relatively smaller differences between the sum of the neutrino masses with and without the light WIMP ($1 \leq \Sigma\,m_{\nu}'/\Sigma\,m_{\nu} \lsim 1.6$).  The neutrino mass ratios, $\Sigma\,m_{\nu}'/\Sigma\,m_{\nu}$, are shown as functions of the light WIMP mass in Fig.\,\ref{fig:mnuvsm} for Majorana and Dirac fermion WIMPs as well as for a scalar WIMP.

\section{Dark Radiation Without Dark Radiation: ``Truly Weak" Light WIMPs}
\label{weak wimp}

\begin{figure}[!t]
\begin{center}
\includegraphics[width=0.60\columnwidth]{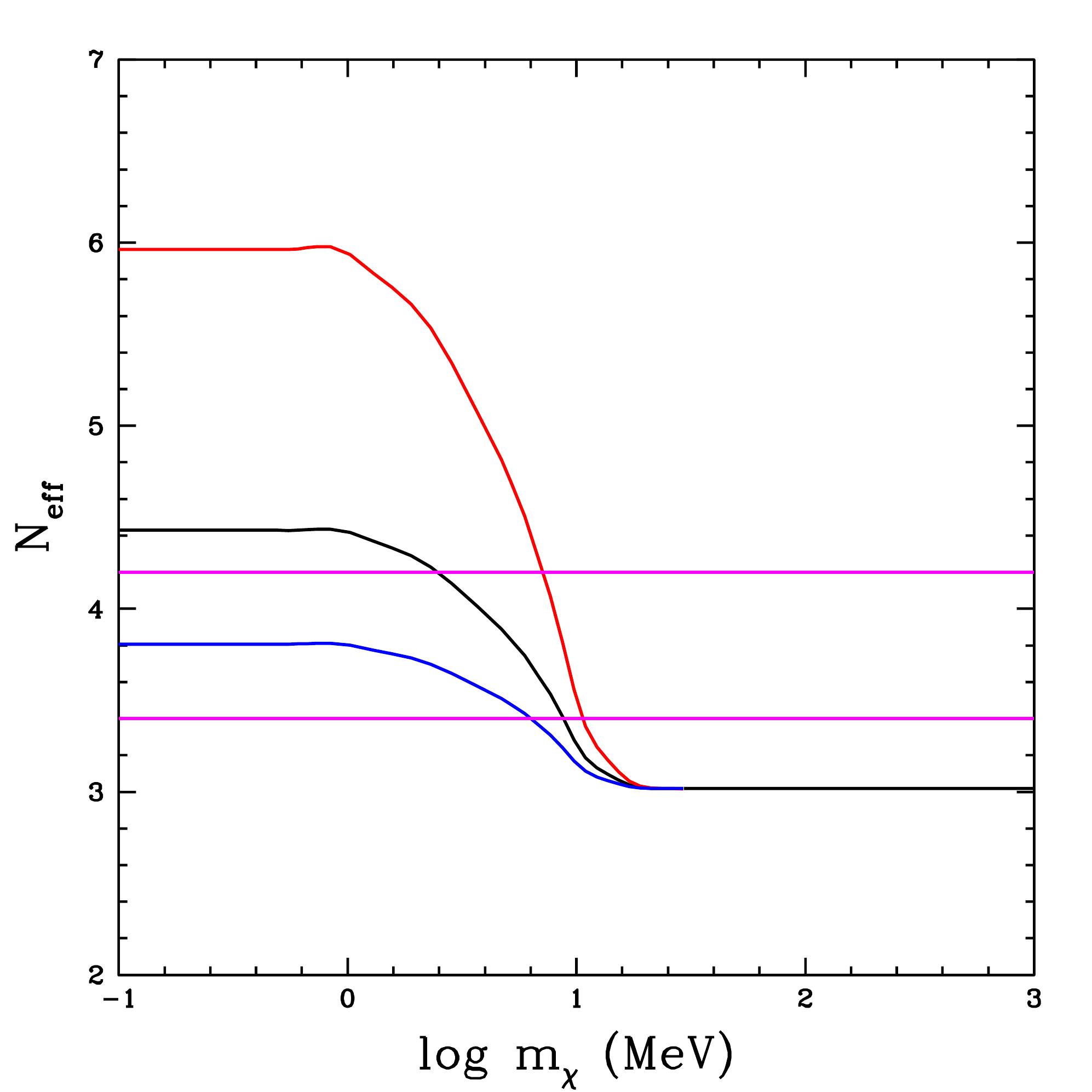}
\caption{The effective number of neutrinos, N$_{\rm eff}$, as a function of the WIMP mass for ``truly weak" WIMPs that only couple to the standard model neutrinos, but not to the photons or the \epm pairs.  The black curve is for a Majorana WIMP, the red curve is for a Dirac WIMP, and the blue curve is for a scalar WIMP.  The horizontal band (purple) corresponds to the $\pm 1\,\sigma$ band consistent with the WMAP9 data.}
\label{fig:neffvsmc}
\end{center}
\end{figure}

As a novel alternative to the case discussed above in \S\,\ref{light wimp}, consider the consequences of a ``truly weak" light WIMP that couples only to the standard model neutrinos, but not to the photons or the \epm pairs \cite{ktw,serpico,boehm2012}.  Assume there are no equivalent neutrinos (\Deln~= 0).  In this case the WIMP annihilation heats the neutrinos (but not the photons), while the annihilation of the \epm pairs heats the photons (but not the decoupled neutrinos).  After the both the \epm pairs and the light WIMPs have annihilated ($T_{\gamma} \rightarrow T_{\gamma 0}$) the ratio of neutrino to photon temperatures can be found by considerations of entropy conservation.  In this case, the entropies (in a comoving volume) of the photons and \epm pairs ($S_{\gamma e}$) and of the neutrinos and the WIMPs ($S_{\nu\chi}$), are conserved individually.  As a result,
\beq
\bigg({T_{\nu} \over T_{\gamma}}\bigg)_{0}^{3} = {1 + 4\tilde{g}_{\chi}\,\phi_{\chi d}/21 \over 1 + 7\,\phi_{ed}/4} = {1 + 4\tilde{g}_{\chi}\,\phi_{\chi d}/21 \over 2.738}\,,\ \,\ \ {11 \over 4}\bigg({T_{\nu} \over T_{\gamma}}\bigg)_{0}^{3} = 1.0045\bigg(1 + {4\tilde{g}_{\chi}\,\phi_{\chi d} \over 21}\bigg)\,.
\eeq
For sufficiently massive WIMPs, for which $\phi_{\chi d} \rightarrow 0$, the usual result, N$_{\rm eff} = 3.018$, is recovered.  But for very light WIMPs, for which $\phi_{\chi d} \rightarrow 1$,
\beq
{11 \over 4}\bigg({T_{\nu} \over T_{\gamma}}\bigg)_{0}^{3} = 1.0045\bigg(1 + {4\,\tilde{g}_{\chi} \over 21}\bigg)\,,\ \,\ \ {\rm N}_{\rm eff} = \neff^{0} = 3.018\bigg(1 + {4\,\tilde{g}_{\chi} \over 21}\bigg)^{4/3}\,.
\eeq
The effective number of neutrinos is shown as a function of the WIMP mass for a truly weak WIMP in Fig.\,\ref{fig:neffvsmc} for Majorana and Dirac WIMPs as well as for a scalar WIMP.  Also shown in Fig.\,\ref{fig:neffvsmc} is the $\pm 1\,\sigma$ band consistent with the WMAP9 value of $\neff$ \cite{hinshaw}.  

The current, pre-Planck CMB estimates suggesting that $\neff > 3$ \cite{hinshaw,melchiorri,hou,calabrese}, are not inconsistent with the absence of equivalent neutrinos (\Deln~= 0).  In contrast to the ``standard" WIMP case, for light WIMPs that couple only to neutrinos, an observational determination of N$_{\rm eff} > 3$ could lead to the mistaken conclusion that \Deln~$> 0$, even in the absence of dark radiation or equivalent neutrinos:\,``Dark radiation without dark radiation".

\subsection{Neutrino Masses In The Presence Of A Truly Weak Light WIMP}

\begin{figure}[!t]
\begin{center}
\includegraphics[width=0.60\columnwidth]{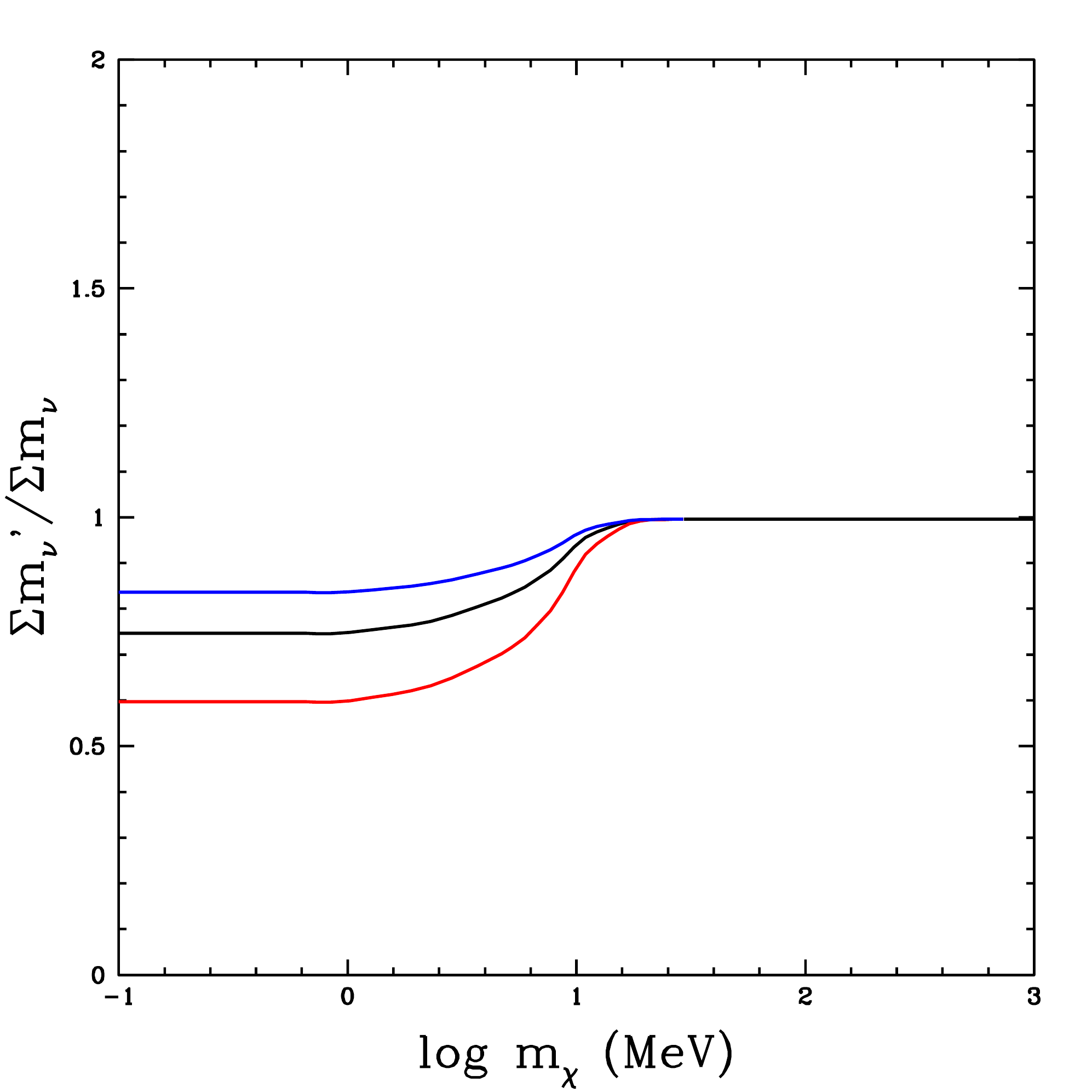}
\caption{As in Fig.\,\ref{fig:mnuvsm}, but for WIMPs that couple only to the SM neutrinos, not to photons.}
\label{fig:mnuvsmb}
\end{center}
\end{figure}

Here, too (see \S\,\ref{numass1}\,,\,\S\,\ref{numass2}\,,\,\S\,\ref{numass3}), the constraint on the sum of the neutrino masses is modified by the presence of a light WIMP that only couples to neutrinos and not to photons.  In this case,
\beq
{\Sigma\,m_{\nu}' \over \Sigma\,m_{\nu}} = \bigg({3 \over N_{\rm eff}}\bigg)^{3/4} \lsim 1\,,
\eeq
tightening the constraint on the sum of the neutrino masses.

Since N$_{\rm eff} > 3$ for truly weak light WIMPs, their presence isn't favorable for the existence of sterile neutrinos.  However, if sterile neutrinos are present, $\Sigma\,m_{\nu}'$ is the sum of the SM and sterile neutrino masses.  As for the case of the ``normally" coupled light WIMPs (\S\,\ref{numass3}), the deviation of N$_{\rm eff}$ from 3 is not very large, resulting in relatively smaller differences between the sum of the neutrino masses with and without the light WIMP ($0.6 \lsim \Sigma\,m_{\nu}'/\Sigma\,m_{\nu} \leq 1$).  $\Sigma\,m_{\nu}'/\Sigma\,m_{\nu}$ is shown as a function of the truly weak, light WIMP mass in Fig.\,\ref{fig:mnuvsmb}.

\section{Summary And Conclusions}
\label{summary}

At late times in the early, radiation dominated Universe, after all the SM particles and light WIMPs ($\chi$), if present, have annihilated, the energy density consists of the contributions from the photons ($\gamma$) and the three SM neutrinos ($\nu$), possibly supplemented by the contribution from \Deln~equivalent neutrinos ($\xi$).  At these late times ($T_{\gamma 0} \ll {\rm min}\{m_{e},m_{\chi}\}$) the ratio, by number, of one species of SM neutrino to the photons is,
\beq
\bigg({n_{\nu} \over n_{\gamma}}\bigg)_{0} = {3 \over 4}\bigg({T_{\nu} \over T_{\gamma}}\bigg)^{3}_{0} = {3 \over 11}\bigg[{11 \over 4}\bigg({T_{\nu} \over T_{\gamma}}\bigg)^{3}_{0}\bigg]\,.
\eeq
Since at least some of the SM neutrinos are sufficiently massive to be non-relativistic at present, the neutrino contribution to the present Universe mass density is $\rho_{\nu 0} = \Sigma\,m_{\nu}n_{\nu 0}$.  In the absence of light WIMPs and equivalent neutrinos, $\Sigma\,m_{\nu} = 94.12\,\Omega_{\nu}h^{2}$, where CMB and LSS data bound the neutrino mass density, $\Omega_{\nu} \leq \Omega_{\rm HDM}$.  An observational constraint on $\Omega_{\rm HDM}h^{2}$ leads to an upper bound to $\Sigma\,m_{\nu}$.  Since the cosmological constraint on the sum of the neutrino masses depends on the frozen out ratio of the number densities of neutrinos to photons, in the more general cases allowing for light WIMPs and equivalent neutrinos the cosmological constraint on the sum of the neutrino masses is modified.  That is, the simple numerical factor connecting $\Sigma\,m_{\nu}$ and $\Omega_{\nu}h^{2}$ is modified, with the conversion factor now depending on the properties of the WIMP and the equivalent neutrinos.  In the presence of a WIMP the late time ratio of SM neutrino and photon temperatures $(T_{\nu}/T_{\gamma})_{0}$ depends on the WIMP mass ($m_{\chi}$) as well as on the SM and equivalent neutrino decoupling temperatures ($T_{\nu d}$ and $T_{\xi d}$, respectively), while the late time ratio of the equivalent neutrino and SM neutrino temperatures $(T_{\xi}/T_{\nu})_{0}$ depends on the equivalent neutrino decoupling temperature.  As a result, $(n_{\nu}/n_{\gamma})_{0}$ is a function of $T_{\xi d}$ and $m_{\chi}$ (and of $T_{\nu d}$).

The neutrino (SM and equivalent neutrinos) contributions to the late time radiation energy density are measured by the effective number of neutrinos, $\neff$,
\beq
{\rm N}_{\rm eff} \equiv \bigg[{11 \over 4}\bigg({T_{\nu} \over T_{\gamma}}\bigg)^{3}_{0}\bigg]^{4/3}\bigg[3 + \Delta{\rm N}_{\nu}\bigg({T_{\xi} \over T_{\nu}}\bigg)^{4}_{0}\bigg] \equiv {\rm N}^{0}_{\rm eff}\bigg(1 + {\Delta{\rm N}^{*}_{\nu} \over 3}\bigg)\,.
\eeq
In general, $\neff$ is  function of \Deln, $T_{\nu d}$, $T_{\xi d}$, and $m_{\chi}$, leading to degeneracies among them for any observationally determined value of $\neff$.

In \S\,\ref{sm} the standard, textbook discussion of neutrino decoupling (freeze out) in the early Universe (no equivalent neutrinos (\Deln~= 0), no light WIMPs) was reviewed, noting how $(n_{\nu}/n_{\gamma})_{0}$ and $\neff$ depend on the choice of the SM neutrino decoupling temperature.  As may be seen in Figs.\,\ref{fig:neffvstnud1}\,-\,\ref{fig:mnuvstnud1}, the earlier the neutrinos decouple (the weaker the weak interactions) the cooler they are relative to the photons and the smaller are $(n_{\nu}/n_{\gamma})_{0}$ (allowing for larger neutrino masses) and N$_{\rm eff}$.  Conversely, the stronger the weak interactions, the later the neutrinos decouple and the larger are the frozen out values of $(n_{\nu}/n_{\gamma})_{0}$ and N$_{\rm eff}$.  As the discussion in \S\,\ref{sm}, and Fig.\,\ref{fig:neffvstnud1} in particular show, if the neutrino decoupling temperature were a free parameter, allowed to vary from $T_{\nu d} \gg m_{t}$ to $T_{\nu d} \ll m_{e}$, the frozen out ratio of neutrinos (one species) to photons would vary by a factor of $\sim 50$, from $(n_{\nu}/n_{\gamma})_{0}\sim 0.015$ to $(n_{\nu}/n_{\gamma})_{0}\sim 0.75$.  The effect on the neutrino mass constraint of this variation in the abundance of neutrinos relative to photons is shown in Fig.\,\ref{fig:mnuvstnud1}.  Allowing $T_{\nu d}$ to be a free parameter, the effective number of neutrinos could assume any value from N$_{\rm eff} \sim 0.06$ to N$_{\rm eff} \sim 11.56$ (see Fig.\,\ref{fig:neffvstnud1}).  In reality, the neutrino decoupling temperature is determined empirically to be $T_{\nu d} \approx 2\,{\rm MeV}$ \cite{enqvist,dolgov,hannestad}.  In the standard, textbook analyses some simplifying assumptions are made (instantaneous decoupling; massless electrons), leading to $(T_{\nu}/T_{\gamma})^{3}_{0} = 4/11$, so that $(n_{\nu}/n_{\gamma})_{0} = 3/11$ and ${\rm N}_{\rm eff} = 3$.   However, in \S\,\ref{sm} it was noted that for the best estimate of $T_{\nu d}$ and assuming the neutrinos decouple instantaneously, there is a small difference from the canonical results (see Fig.\,\ref{fig:neffvstnud2}); $(T_{\nu}/T_{\gamma})^{3}_{0} \rightarrow 1.006(4/11)$, so that $(n_{\nu}/n_{\gamma})_{0} \rightarrow 1.006(3/11)$ and N$_{\rm eff} \rightarrow 3.018$.

With these results as prologue, in \S\,\ref{equivnus} the standard model of particle physics was extended to allow for the presence of \Deln~equivalent neutrinos ($\xi$).  Fixing the SM neutrino decoupling temperature at $T_{\nu d} = 2\,{\rm MeV}$, the connection between the equivalent neutrino decoupling temperature ($T_{\xi d}$) and the late time ratio of the SM neutrino to photon temperatures was explored along with the changes to the corresponding values of $(n_{\nu}/n_{\gamma})_{0}$ (and its implication for the constraint on the sum of the neutrino masses) and N$_{\rm eff}$ (see Figs.\,\ref{fig:neffvstd}\,-\,\ref{fig:mvstd}).  As may be seen in Fig.\,\ref{fig:neffvstd}, depending on when an equivalent neutrino decouples (how weakly it interacts with the SM particles), one equivalent neutrino (\ie, a very light, Majorana fermion) need not contribute \Deln~= 1 to N$_{\rm eff}$.  In the equivalent neutrino contribution to $\neff$ there is a degeneracy between \Deln~and $T_{\xi d}$.  As $T_{\xi d}$ decreases from $\gg m_{t}$ to $\ll m_{e}$, the contribution to N$_{\rm eff}$ from one equivalent neutrino increases from 0.05 to 3.85, while the contribution from the SM neutrinos increases from 3.02 to 4.82, and N$_{\rm eff}$ increases from 3.07 to 8.67 (7.02 for a scalar equivalent neutrino); see Figs.\,\ref{fig:neffvstd}\,-\,\ref{fig:neffvstdc}.  As noted in \S\,\ref{equivnus}, sterile neutrinos are a special case of the more general equivalent neutrinos.  Sterile neutrinos, very light Majorana fermions that decouple along with the SM neutrinos ($T_{\xi d} = T_{\nu d} = 2\,{\rm MeV}$), simplify the connection between $\neff$ and \Deln~by eliminating the degeneracy between \Deln~and $T_{\xi d}$.  In this case, N$_{\rm eff} = 3.018(1 + \Delta{\rm N}_{\nu}/3)$, corresponding to N$_{\rm eff} = 4.02\,(5.03)$ for one (two) sterile neutrinos.

In \S\,\ref{light wimp} and \S\,\ref{weak wimp} the connections between a light WIMP and equivalent neutrinos were explored.  A relatively light WIMP, whether or not it qualifies as a dark matter candidate, will annihilate late during the early evolution of the Universe, heating the SM particles, including possibly the neutrinos (SM and equivalent).  For a sufficiently light WIMP ($m_{\chi} \lsim 20\,{\rm MeV}$) without enhanced coupling to the SM neutrinos, late time annihilation may heat the photons relative to the decoupled neutrinos, reducing $(T_{\nu}/T_{\gamma})_{0}$ below what it would be in the absence of the WIMP, resulting in N$_{\rm eff} < 3$, even in the presence of the three SM neutrinos.  This allows for additional equivalent neutrinos, \Deln~$> 0$, even if observations should determine that N$_{\rm eff} \approx 3$ \cite{hoscherrer1,hoscherrer2}.  Indeed, as found in \S\,\ref{light wimp} and as shown in Fig.\,\ref{fig:neffvsma}, for a very light ($m_{\chi} \lsim m_{e}$) Majorana fermion WIMP, N$_{\rm eff} \rightarrow 2.09$, allowing for the consistency of one or even two sterile neutrinos with current CMB constraints \cite{hinshaw,melchiorri,calabrese} (see Fig.\,\ref{fig:neffvsm3}).  As noted in \S\,\ref{degeneracy} and illustrated in Figs.\,\ref{fig:delnvsm}\,\,-\,\ref{fig:delnvsma}, in the presence of a light WIMP there is a degeneracy between the WIMP mass and the combination $\Delta{\rm N}_{\nu}^{*} \equiv \Delta{\rm N}_{\nu}(T_{\xi}/T_{\nu})^{4}_{0}$.  Depending on the observationally determined value of N$_{\rm eff}$, there may be several combinations of $m_{\chi}$, $\Delta{\rm N}_{\nu}$, and $T_{\xi d}$ that are consistent with the same value of N$_{\rm eff}$.  To illustrate this point the case of three right-handed neutrinos\,\cite{ags}, where \Deln~= 3 and $\Delta{\rm N}_{\nu}^{*} = 3(T_{\xi}/T_{\nu})^{4}_{0}$, was revisited.  It was noted that for $m_{\chi} \lsim 1\,{\rm MeV}$ and $\neff = 3$, $\Delta{\rm N}_{\nu}^{*} \approx 1.3$, requiring $(T_{\xi}/T_{\nu})^{4}_{0} \approx 0.4$, which can be achieved provided that $T_{\xi d} \approx 120\,{\rm MeV} \approx 60\,T_{\nu d}$ \cite{ags}.  Such a high decoupling temperature for the three right-handed neutrinos could result from their being coupled to a heavier Z boson, $M_{Z'}/M_{Z} \approx 8$ or $M_{Z'} \approx 0.7\,{\rm TeV}$.  In general, in the presence of a sufficiently light WIMP, $\neff = 3$ is no guarantee of the absence of equivalent neutrinos.

The discussion in \S\,\ref{weak wimp} considered the effects of a ``truly weak" light WIMP, a particle that couples only to the SM neutrinos but not to the other SM particles (in particular, it does not couple to the photons and the \epm pairs) \cite{ktw,serpico,boehm2012}.  Before the SM neutrinos decouple ($T_{\gamma} \geq T_{\nu d}$), $T_{\nu} = T_{\gamma}$.  However, when $T_{\gamma} < T_{\nu d}$ \epm annihilation heats the photons but not the decoupled SM neutrinos.  In contrast, when the truly weak WIMP annihilates it heats the SM neutrinos but not the photons, bringing the late time neutrino and photon temperatures closer together.  As was the case in \S\,\ref{light wimp}, the simple connection between N$_{\rm eff}$ and \Deln~is broken.  In the presence of a truly weak, light WIMP it is possible to have N$_{\rm eff} > 3$ even if \Deln~= 0\,: Dark radiation without dark radiation.  

As the key points presented here have shown, there's more to a measurement of the effective number of neutrinos than meets the eye, at least at first sight.  In the presence of a sufficiently light WIMP, $\neff$ depends on the WIMP mass, $m_{\chi}$, as well as its nature (fermion or boson) and its coupling, or not, to the SM neutrinos, on the number of equivalent neutrinos, \Deln~(and on their nature as well), and on the equivalent neutrino decoupling temperature, $T_{\xi d}$.  A measurement of $\neff = 3$, within the observational uncertainties, is not evidence for the absence of equivalent neutrinos, sterile or otherwise.  Conversely, a measurement of $\neff > 3$, accounting for the observational uncertainties, does not, by itself, establish the presence of equivalent neutrinos or dark radiation.

\section*{Acknowledgments}
I thank L. Anchordoqui, A. Evrard, H. Goldberg, G. Hinshaw, C. McCabe, L. Page, E. Rozo, and R. J. Scherrer for helpful discussions.  Special thanks are due J. Beacom for helpful discussions and for careful readings of earlier versions of this manuscript.  This research was supported by DOE Grant DE-FG02-91ER40690.  



\begin{thebibliography}{99}

\bibitem{gross}
D.~J.~Gross and F.~Wilczek,
Phys.\ Rev.\ Lett. {\bf 30} (1973) 1343.

\bibitem{politzer}
H.~D.~Politzer,
Phys.\ Rev.\ Lett. {\bf 30} (1973) 1346.

\bibitem{perl}
M.~L.~Perl, {\it et al.},
Phys.\ Lett.\ B {\bf 63} (1976) 466.

\bibitem{ssg}
G.~Steigman, D.~N.~Schramm, and J.~E.~Gunn,
Phys.\ Lett.\ B {\bf 66} (1977) 202.

\bibitem{hoyle}
F.~Hoyle and R.~J.~Tayler,
Nature {\bf 203} (1964) 1108.

\bibitem{pje}
P.~J.~E.~Peebles,
Phys.\ Rev.\ Lett.\ {\bf 16} (1966) 410.

\bibitem{shvartsman}
V.~F.~Shvartsman,
JETP\ Lett.\ {\bf 9} (1969) 184.

\bibitem{steigman}
G.~Steigman,
Advances in High Energy Phys. {\bf 2012} (2012) 268321 [arXiv:1208.0186 [hep-ph]].

\bibitem{komatsu}
E.~Komatsu, {\it et al.},
Ap.\ J.\ S\ {\bf 192} (2011) 18.

\bibitem{dunkley}
J.~Dunkley, {\it et al.},
Ap.\ J.\ {\bf 739} (2011) 52.

\bibitem{keisler}
R.~Keisler, {\it et al.},
Ap.\ J.\ {\bf 743} (2011) 28.

\bibitem{archidiacono}
M.~Archidiacono, E.~Calabrese, and A.~Melchiorri,
Phys.\ Rev\ D {\bf 84} (2011) 123008.

\bibitem{galli}
S.~Galli, M.~Martinelli, A.~Melchiorri, L.~Pagano, B.~D.~Sherwin, and D.~N.~Spergel,
Phys.\ Rev.\ D {\bf 82} (2010) 123504.

\bibitem{enqvist}
K.~Enqvist, K.~Kainulainen, and V.~Semikoz,
Nucl.\ Phys.\ B {\bf 374} (1992) 392.

\bibitem{dolgov}
A.~D.~Dolgov, 
Phys.\ Rept. {\bf 370} (2002), 333.

\bibitem{hannestad}
S.~Hannestad,
Phys.\ Rev.\ D {\bf 65} (2002) 083006.

\bibitem{mangano}
G.~Mangano, G.~Miele, S.~Pastor, T.~Pinto, O.~Pisanti, and P.~D.~Serpico,
Nucl.\ Phys.\ B {\bf 729} (2005) 221.

\bibitem{ags}
L.~Anchordoqui, H.~Goldberg, and G.~Steigman,
Phys.\ Lett. B {\bf 718} (2013) 1162.

\bibitem{hinshaw}
G.~Hinshaw, {\it et al.},
[arXiv:1212.5226 [astro-ph.CO]].

\bibitem{calabrese}
E.~Calabrese, {\it et al.},
[arXiv:1302.1841 [astro-ph.CO]].

\bibitem{hou}
Z.~Hou, {\it al},
[arXiv:1212.6267 [astro-ph.CO]].

\bibitem{melchiorri}
E.~Di Valentino, {\it et al.},
[arXiv:1301.7343 [astro-ph.CO]].

\bibitem{ktw}
E.~Kolb, M.~S.~Turner, and T.~P.~Walker,
Phys.\ Rev.\ D {\bf 34} (1986) 2197.

\bibitem{serpico}
P.~D.~Serpico and G.~G.~Raffelt,
Phys.\ Rev.\ D {\bf 70} (2004) 043526.

\bibitem{boehm1}
C.~Boehm, T.~A.~Ensslin, and J.~Silk, 
J.\ Phys.\ G {\bf 30} (2004) 279.

\bibitem{boehm2}
C.~Boehm and P.~Fayet, 
Nucl.\ Phys.\ B {\bf 683} (2004) 219.

\bibitem{boehm3}
C.~Boehm, D.~Hooper, J.~Silk, and M.~Casse, 
Phys.\ Rev.\ Lett. {\bf 92} (2004) 101301.
 
\bibitem{hooper1}
D.~Hooper, F.~Ferrer, C. Boehm, J.~Silk, J.~Paul, N.~W.~Evans, and M.~Casse, 
Phys.\ Rev.\ Lett. {\bf 93} (2004) 161302.

\bibitem{boehm4}
C.~Boehm, P.~Fayet, and J.~Silk, 
Phys.\ Rev.\ D {\bf 69} (2004) 101302.

\bibitem{ahn}
K.~Ahn and E.~Komatsu, 
Phys.\ Rev.\ D {\bf 72} (2005) 061301.

\bibitem{fayet}
P.~Fayet, D.~Hooper, and G.~Sigl, 
Phys.\ Rev.\ Lett. {\bf 96} (2006) 211302.

\bibitem{hooper2}
D.~Hooper, M.~Kaplinghat, L.~E.~Strigari, and K.~M.~Zurek, 
Phys.\ Rev.\ D {\bf 76} (2007) 103515.

\bibitem{hooper3}
D.~Hooper and K.~M.~Zurek, 
Phys.\ Rev.\ D {\bf 77} (2008) 087302.

\bibitem{feng}
J.~L.~Feng and J.~Kumar, 
Phys.\ Rev.\ Lett. {\bf 101} (2008) 231301.

\bibitem{hoscherrer1}
C.~M.~Ho and R.~J.~Scherrer,
[arXiv:1208.4347 [astro-ph.CO]]

\bibitem{hoscherrer2}
C.~M.~Ho and R.~J.~Scherrer,
[arXiv:1212.1689 [astro-ph.CO]]

\bibitem{boehm2012}
C.~Boehm, M.~J.~Dolan, and C.~McCabe,
[arXiv:1207.0497 [astro-ph.CO]]

\end{thebibliography}
\end{document}